\documentclass{jfm}
\usepackage{graphicx}
\usepackage{float}
\usepackage{rotating}
\usepackage{multirow}
\usepackage{amsfonts,amssymb,latexsym,amsmath,pifont,mathdots,mathtools,natbib}
\usepackage{color}
\usepackage{setspace}
\usepackage{subfig}
\usepackage{url}
\usepackage{epstopdf,epsfig}
\newcommand{\red}[1]{{{\color{black}#1}}}
\title{Mean turbulence statistics in boundary layers over high-porosity foams}
\shorttitle{Turbulent boundary layers over high-porosity foams}

\author{Christoph Efstathiou$^{1}$ \and Mitul Luhar$^{1}$\thanks{Email address for correspondence: luhar@usc.edu}}
\affiliation{$^{1}$Department of Aerospace and Mechanical Engineering, University of Southern California, Los Angeles, CA 90089}
\shortauthor{C. Efstathiou and M. Luhar}

\begin{document}
\maketitle

\begin{abstract}
This paper reports turbulent boundary layer measurements made over open-cell reticulated foams with varying pore size \red{and thickness, but constant porosity ($\epsilon \approx 0.97$)}. The foams were flush-mounted into a cutout on a flat plate.  A Laser Doppler Velocimeter (LDV) was used to measure mean streamwise velocity and turbulence intensity immediately upstream of the porous section, and at \red{multiple} measurement stations along the porous substrate. The friction Reynolds number upstream of the porous section was $Re_\tau \approx 1690$. For all \red{but the thickest} foam tested, the internal boundary layer was fully developed by $<10 \delta$ downstream from the porous transition, where $\delta$ is the boundary layer thickness. Fully developed mean velocity profiles showed the presence of a substantial slip velocity at the porous interface ($>30\%$ of the free stream velocity) and a mean velocity deficit relative to the canonical smooth-wall profile further from the wall. While the magnitude of the mean velocity deficit increased with average pore size, the slip velocity remained approximately constant.  Fits to the mean velocity profile suggest that the logarithmic region is shifted relative to a smooth wall, and that this shift increases with pore size until it becomes comparable to substrate thickness $h$. For all foams, the turbulence intensity was found to be elevated further into the boundary layer to $y/ \delta \approx 0.2$.  An outer peak in intensity was also evident for the largest pore sizes.  Velocity spectra indicate that this outer peak is associated with large-scale structures resembling Kelvin-Helmholtz vortices that have streamwise length scale $2\delta-4\delta$.  Skewness profiles suggest that these large-scale structures may have an amplitude-modulating effect on the interfacial turbulence.
\end{abstract}


\section{Introduction}\label{sec:intro}

Turbulent flows of scientific and engineering interest are often bounded by walls that are not smooth, solid, or uniform. Manufacturing techniques, operational requirements and natural evolution often lead to non-uniform, rough, and porous boundaries. Examples include flows over heat exchangers, forest canopies, bird feathers and river beds \citep[e.g.,][]{finnigan2000turbulence,jimenez2004turbulent,ghisalberti2009obstructed,favier2009passive,manes2011turbulent,jaworski2013aerodynamic,chandesris2013direct,kim2016experimental}.  Porous boundaries also enable active flow control through suction and blowing for use in drag reduction and delaying transition from laminar to turbulent flow \citep[e.g.,][]{parikh2011passive}.

Despite the potential applications, relatively little is known about the relationship between turbulent flows and porous substrates.  For example, it is unclear how established features of smooth-wall flows, such as the self-sustaining near-wall cycle, the logarithmic region in the mean profile, the larger-scale structures found further from the wall, and the interaction between the inner and outer regions of the flow \citep[e.g. the amplitude modulation phenomenon;][]{marusic2010predictive} are modified over porous surfaces. As a result, there are few models that can predict how a porous substrate of known geometry will influence the mean flow field and turbulent statistics. A key limitation is that few experimental and numerical datasets exist for turbulent flows over porous media.  Further, most previous experimental datasets (motivated primarily by flows over packed sediment beds or canopies) include limited near-wall measurements, while previous numerical simulations have been restricted to relatively low Reynolds numbers. \red{In addition, almost all previous studies on turbulent flow over porous substrates have employed relatively thick media, such that flow penetration into the substrate depends only on pore size or permeability \citep[see e.g.,][]{manes2011turbulent}. However, in many natural and engineered systems, the porous substrate can be of finite thickness and bounded by a solid boundary. Examples include feathers or fur in natural locomotion \citep{itoh2006turbulent,jaworski2013aerodynamic}, and heat exchangers employing metal foams \citep{mahjoob2008synthesis}. In such systems, substrate thickness can also have an important influence on flow development and the eventual equilibrium state. At the limit where porous medium thickness becomes comparable to pore size, the substrate can essentially be considered a rough wall. The transition from this rough-wall limit to more typical porous medium behaviour is not fully understood.  The experimental study described in this paper seeks to address some of the limitations described above.}


\subsection{Previous studies}\label{sec:literature}
Previous numerical efforts investigating the effect of porous boundaries include the early direct numerical simulations (DNS) performed by \citet{jimenez2001turbulent}, which employed an effective admittance coefficient linking the wall-pressure and wall-normal velocity to model the porous wall.  More recent DNS efforts have either employed the volume-averaged Navier-Stokes equations, where the porous substrate is modeled as a resistive medium \citep{breugem2006influence}, or explicitly modeled the porous medium as an array of cubes \citep{chandesris2013direct}. 

In particular, the results of \citet{breugem2006influence} showed important deviations from turbulent channel flow over smooth walls.  The mean velocity was significantly reduced across much of the channel for the cases with high porosity ($\epsilon = 0.8$ and $0.95$) and this was accompanied by a skin friction coefficient increase of up to $30\%$.  The presence of the porous substrate also led to substantial changes in flow structure.  Large spanwise rollers with streamwise length scale comparable to the channel height were observed at the porous interface.  Given the presence of an inflection point in the mean velocity profile near the porous interface, \citet{breugem2006influence} attributed the emergence of these large-scale structures to a Kelvin-Helmholtz type of instability mechanism \citep[see also][]{jimenez2001turbulent,chandesris2013direct}. In addition, the near-wall cycle comprising streaks and streamwise vortices, a staple of smooth wall turbulent boundary layers, was substantially weakened over the porous substrate.  This weakening was linked to a reduction in the so-called wall-blocking effect and enhanced turbulent transport across the interface. \citet{breugem2006influence} also found that the root mean square (rms) of the wall-normal velocity fluctuations did not exhibit outer layer similarity and suggested two potential causes for this (i) more vigorous sweeps and ejections due to the absence of an impermeable wall and (ii) insufficient scale separation between the channel half-height and the penetration distance into the porous substrate. \citet{chandesris2013direct} noted broadly similar trends in their DNS, which also considered thermal transport.

\red{\citet{rosti2015direct} performed DNS studies of turbulent channel flow at $Re_\tau = 180$ using a VANS formulation that allowed for the porosity and permeability to be decoupled. These simulations showed that even relatively low wall permeabilities led to substantial modification of the turbulent flow in the open channel. Further, despite a substantial variation in the porosities tested ($\epsilon = 0.3-0.9$), the flow was found to be much more sensitive to permeability. \citet{motlagh2016pod} performed a VANS Large Eddy Simulation (LES) of channel flow over porous substrates with $\epsilon=0, 0.8$ and $0.95$ at bulk Reynolds number 5500. Based on proper orthogonal decomposition (POD) of the flow field, they showed that large-scale features prevalent in smooth-walled flows were still present but beginning to break down for the lower porosity case. Consistent with the results of \citet{breugem2006influence}, for the higher porosity case, these structures were replaced by spanwise-elongated counter-rotating vortices.}

\red{Recently, \citet{kuwata2016lattice} and \citet{kuwata2017direct} have performed Lattice-Boltzman simulations over rough walls, staggered arrays of cubes, as well as anisotropic porous media.  In particular, \citet{kuwata2017direct} considered a model system in which the streamwise, spanwise, and wall-normal permeabilities could be altered individually. These simulations showed that the streamwise permeability is instrumental in preventing the high- and low-speed streaks associated with the near-wall cycle. The simulations also indicated that, unlike the streamwise and spanwise permeabilities, the wall-normal permeability does not significantly enhance turbulence intensities.}

Similar to turbulent flows over smooth and rough walls, previous studies suggest that a logarithmic region in the mean profile $U(y)$ can also be expected in turbulent flows over porous media.  This logarithmic region is often parametrized as:
\begin{equation}\label{eq:log-law}
U^+ = \frac{1}{\kappa}  \ln{\left(\frac{y + y_d}{k_0} \right)} = \frac{1}{\kappa} \ln{(y^+ + y_d^+)} + B -\Delta U^+,
\end{equation}
where $\kappa$ and $B$ are the von Karman and additive constants, $y$ is the wall-normal distance from the porous interface, $y_d$ is the shift of the logarithmic layer (or zero-plane displacement height), $k_0$ is the equivalent roughness height, and $\Delta U^+$ is the roughness function \citep{jimenez2004turbulent}. A superscript $+$ indicates quantities normalized by the friction velocity $u_\tau$  and the kinematic viscosity $\nu$.  Fits to the mean velocity profiles obtained in DNS by \citet{breugem2006influence} suggest that the von Karman constant decreases from $\kappa \approx 0.4$ for the smooth wall case to $\kappa = 0.23$ for the most porous substrate.  However, further tests were recommended at higher Reynolds number to confirm this effect.

Experimental efforts in this realm have considered flows over beds of packed spheres, perforated sheets, foams, as well as seal fur \citep[e.g.,][]{ruff1972turbulent,zagni1976channel,kong1982turbulent,itoh2006turbulent,suga2010effects,manes2011turbulent,kim2016experimental}. Interestingly, the seal fur experiments show a reduction in skin friction though the exact mechanism behind this remains to be understood \citep{itoh2006turbulent}. \citet{kong1982turbulent} investigated the effect of small scale roughness over smooth, rough, and porous surfaces on turbulent boundary layers over bluff bodies. The porous boundaries consisted of perforated sheets and mesh screens.  These experiments showed that the turbulent Reynolds stresses increased near the interface, as did the skin friction. However, it is difficult to separate roughness effects from permeability effects for these experiments.  The mean velocity profile was shifted by $\Delta U^+ \approx 3-4$, which was similar to the shift obtained over an impermeable rough wall of similar geometry. 

\citet{suga2010effects} studied laminar and turbulent channel flow over foamed ceramics with porosity $\epsilon \approx 0.8$ and varying pore sizes via Particle Image Velocimetry (PIV).  These experiments were carried out at relatively low Reynolds numbers (bulk Reynolds number $Re_b \le 10,200$).  The measurements indicated that the transition to turbulence occurs at lower Reynolds number over the porous media.  Further, turbulence intensities were generally enhanced over the porous medium, and the displacement and roughness heights in the modified logarithmic law were found to increase with increasing pore size and permeability. \red{\citet{detert2010synoptic} also used PIV to study the flow over packed spheres and gravel from the river Rhine in an open channel flow with friction Reynolds numbers $Re_\delta = 1.88-14.7\times 10^3$. At the relatively low porosities tested ($\epsilon = 0.26-0.33$), the flow exhibited many of the large-scale structures observed in turbulent boundary layers over smooth walls, including hairpin vortex packages. The observed flow patterns were also found to be relatively insensitive to Reynolds number.}

\cite{manes2011turbulent} made open channel turbulent flow measurements over very porous ($\epsilon > 0.96$) polyurethane foam mattresses with 10-60 pores per inch (ppi) at Reynolds number $Re_\tau>2000$ using an LDV. In contrast to \cite{breugem2006influence}, the results obtained by \citet{manes2011turbulent} supported the outer layer similarity hypothesis.  This suggests that the lack of outer-layer similarity in the DNS carried out by \citet{breugem2006influence} resulted from insufficient scale separation between the inner and outer layers of the flow, which is analogous to the breakdown of outer-layer similarity in shallow boundary layers over rough walls \citep{jimenez2004turbulent}. Further, \citet{manes2011turbulent} reported a reduction in the streamwise fluctuation intensity near the wall and an increase in the intensity of wall-normal fluctuations, which they attributed to reduced wall blocking.  For the logarithmic region in the mean profile, a fitting procedure similar to the one employed by \citet{breugem2006influence} and \citet{suga2010effects} yielded $\kappa \approx 0.3$ over the porous media and an equivalent roughness height $k_0$ that generally increased with increasing pore size.  However, the mean profiles did not exhibit a clear logarithmic region over the 10 ppi foam.  Since the flow is expected to penetrate further into the foam as pore size increases, \citet{manes2011turbulent} suggested that a lack of scale separation between the penetration distance and the water depth may have influenced the results.

Finally, note that there are substantial similarities in turbulent flows over porous media and vegetation or urban canopies \citep[e.g.,][]{finnigan2000turbulence,poggi2004effect,white2007shear}.  For example, large-scale structures resembling Kelvin-Helmholtz vortices play an essential role in dictating mass and momentum transport at the interface \citep[see e.g.,][]{finnigan2000turbulence} and a shifted log law of the form shown in equation (\ref{eq:log-law}) provides a reasonable fit for the velocity profile above the canopy. In a comparative study of obstructed shear flows, \citet{ghisalberti2009obstructed} suggested that an inflection point exists in the mean profile if the distance to which the flow penetrates into the porous medium or canopy is much smaller than the height of the medium.  This penetration distance is expected to scale as $\sqrt{k}$ for porous substrates \citep[e.g.,][]{battiato2012self}, where $k$ is the permeability, and $(C_D a)^{-1}$ for canopies where, $C_D$ is a drag coefficient and $a$ is the frontal area per unit volume of the canopy.  For densely packed or tall substrates where an inflection point is observed, the interfacial dynamics are dominated by structures resembling Kelvin-Helmholtz vortices. In such cases, the slip velocity at the interface depends primarily on the friction velocity, with little dependence on substrate geometry.  Further, the interfacial turbulence also tends to be more isotropic, such that intensity of the wall-normal velocity fluctuations is comparable to the intensity of the streamwise fluctuations.

\subsection{Contribution and outline}\label{sec:outline}
The present study builds on the experiments pursued by \citet{manes2011turbulent} to provide further insight into the near-wall flow physics over high-porosity surfaces.  An LDV was used to measure streamwise velocity profiles over commercially-available foams with systematically varying pore sizes \red{and thicknesses} at moderately high Reynolds number. For reference, the friction Reynolds number over the smooth wall upstream of the porous section was $Re_\tau = u_\tau \delta/\nu \approx 1690$, where $\delta$ is the $99\%$ boundary layer thickness. The velocity profiles include measurements very close to the porous interface (corresponding to 2-3 viscous units over the smooth wall). \red{Unfortunately, wall-normal velocities were not measured due to instrumentation limitations.}

Morphologically, the foams tested in this study are similar to those considered by \cite{manes2011turbulent}.  However, one important distinction is the thickness of the foam layer, $h$.  In an effort to isolate the effects of pore size, $s$, and permeability, $k$, \citet{manes2011turbulent} considered very thick porous media with $h \gg \sqrt{k}$ and $h \gg s$.  The scale separation is more limited in the present study, with \red{$h/s \approx 4.3-44$ and $h/\sqrt{k} \approx 36-160$}. As a result, porous layer thickness does influence the flow for foams with the largest pore sizes and permeabilities.  In other words, the present study provides insight into the transition between conditions in which the shear flow penetrates across the entire porous domain and conditions in which the shear layer only reaches a small distance into the porous medium. In a sense, this is analogous to the transition between sparse and dense canopy behaviour observed in vegetated shear flows \citep{luhar2008interaction, ghisalberti2009obstructed}.

In recent years, the amplitude modulation phenomenon observed in smooth- and rough-walled turbulent flows has led to a promising class of predictive models \citep[e.g.,][]{mathis2009large,marusic2010predictive,mathis2013estimating,pathikonda2017inner}.  Specifically, it has been observed that the so-called very-large-scale motions (VLSMs) prevalent in the logarithmic region of the flow at high Reynolds number \citep{smits2011high} have a modulating influence on the intensity of the near-wall turbulence.  Further, it has been shown that there is an intrinsic link between this phenomenon and the skewness of the streamwise velocity fluctuations \citep{mathis2011relationship,duvvuri2015triadic}.  Therefore, by considering the skewness of the streamwise velocity, we also evaluate whether such interactions between the inner and outer region persist in turbulent flows over porous media that may be dominated by a different class of large-scale structure, i.e. Kelvin-Helmholtz vortices.

The remainder of this paper is structured as follows: \S\ref{sec:expts} describes the experiments, providing details on the flow facility, porous substrates, and diagnostic techniques; \S\ref{sec:development} shows results on flow development over the porous foams; \S\ref{sec:poresize} illustrates the effect of pore size on mean turbulence statistics; \red{\S\ref{sec:thickness} explores the effect of substrate thickness; \S\ref{sec:spectra} presents velocity spectra over all the different foams tested; }\S\ref{sec:discussion} tests whether a shifted logarithmic region exists over the porous surfaces, evaluates the relative effect of pore size and substrate thickness, and also considers how porous substrates may affect the amplitude modulation phenomenon. Conclusions are presented in \S\ref{sec:conclusions}.

\section{Experimental methods}\label{sec:expts}

\subsection{Flow facility and flat plate apparatus}\label{sec:facility}

\begin{figure}
	\centering\includegraphics[width=\textwidth]{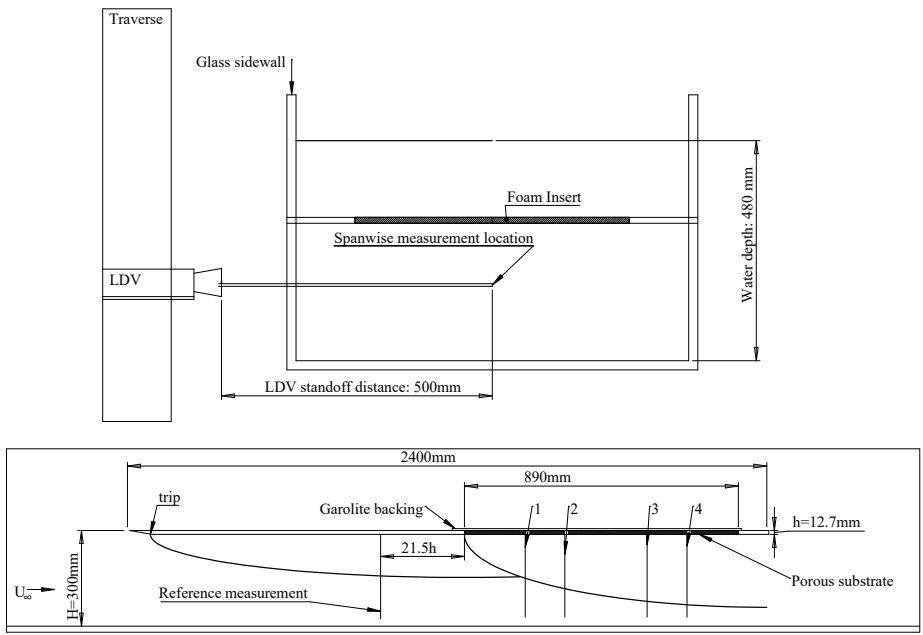}
	\caption{Schematics showing flat plate apparatus in the wall normal-spanwise plane ($y-z$, top) and the streamwise-wall normal plane ($x-y$, bottom).  Measurement positions marked 1-4 were located at $x/h = 11, 21, 42, 53$, where $h=12.7$ mm is the baseline substrate thickness, and $x=0$ is defined as the smooth-porous wall transition. The smooth wall reference measurements were made at a location $x/h=-21.5$.  Note that the wall-normal coordinate is $y$.}
	\label{fig:diag}
\end{figure}

All experiments were conducted in the USC water channel, a free-surface, recirculating facility with glass along both sidewalls and at the bottom to allow for unrestricted optical access.   The water channel has a test section of length 762 cm, width 89 cm, and height 61 cm, and is capable of generating free-stream velocities up to 70 cm/s with background turbulence levels $<1\%$ at a water depth of 48 cm.  The temperature for all experiments was $23\pm 0.5^\circ C$ for which the kinematic viscosity is $\nu = 0.93 \times 10^{-2}$ cm$^2$/s.

Figure \ref{fig:diag} provides a schematic of the experimental setup in the wall normal-spanwise (top) and streamwise-wall normal (bottom) planes.  A 240 cm long flat plate was suspended from precision rails at a height $H=30$ cm above the test section bottom. To avoid free-surface effects, measurements were made below the flat plate. The nominal free-stream velocity was set at $U_e = 58$ cm/s for all the experiments. The confinement between the flat plate and bottom of the channel naturally led to a slightly favorable pressure gradient and slight free stream velocity increase  along the plate $(\leq 3\%)$, however the non-dimensional acceleration parameter, $\Lambda = \frac{\nu}{U_e^2} \frac{dU_e}{dx}$ was on the order of $10^{-7}$ suggesting any pressure gradient effects are likely to be mild \citep{PatelPreston1965,de2000reynolds,schultz2007rough}.

A cutout of length 89 cm and width 60 cm was located 130 cm downstream of the leading edge. Smooth and porous surfaces were substituted into the cutout, flush with the smooth plate around it. The porous test specimens, described in further detail below, were bonded to a solid Garolite\textsuperscript{TM} sheet to provide a rigid structure and prevent bleed through. \red{The setup was designed to accommodate porous substrates of thicknesses $h=6.35, 12.7$ and $25.4$ mm.} Care was taken to minimize gaps and ensure a smooth transition from solid to porous substrate. Velocity profiles were measured over the smooth section upstream of the cutout and at four additional locations over the porous walls.  The flow was tripped by a wire of 0.5 mm diameter located 15 cm downstream of the leading edge.

\subsection{Velocity measurement}\label{sec:diagnostics}
Measurements of \red{streamwise velocity, $u$,} were made at the channel centerline using a Laser Doppler Velocimeter (LDV, MSE Inc.) with a 50 cm standoff distance and a measurement volume of $300 \mu$m by $150 \mu$m by $1000 \mu$ m ($x$ by $y$ by $z$). The LDV was mounted on a precision traverse capable of 16 $\mu$m resolution. Polyamide seeding particles (PSP) with an average size of 5 $\mu$m were used to seed the flow. While the large standoff distance enabled measurements at the channel centerline, it also limited data rates to 50 Hz in the free-stream and less than 1 Hz at the stations closest to the wall.  A minimum of 800 data points in the regions with lowest velocity and 4000-25000 data points in regions of higher velocity were collected to ensure fully converged statistics. In all cases, a preliminary coarse velocity profile was measured to determine the boundary layer thickness and the approximate location of the wall.  This preliminary profile was used to generate a finer logarithmically-spaced vertical grid for the actual measurements.  The nominal smooth-wall location ($y^\prime = 0$) was identified as the position where the data rate dropped to zero. For all the profiles, the vertical grid resolution was reduced to 30 $\mu$m in the near-wall region. Given the 150$\mu$m measurement volume and the $30\mu$m vertical resolution, the nominal estimate for wall-normal location suffers from an uncertainty of $\le 75\mu$m.

\begin{figure}
	\centering\includegraphics[scale=0.8]{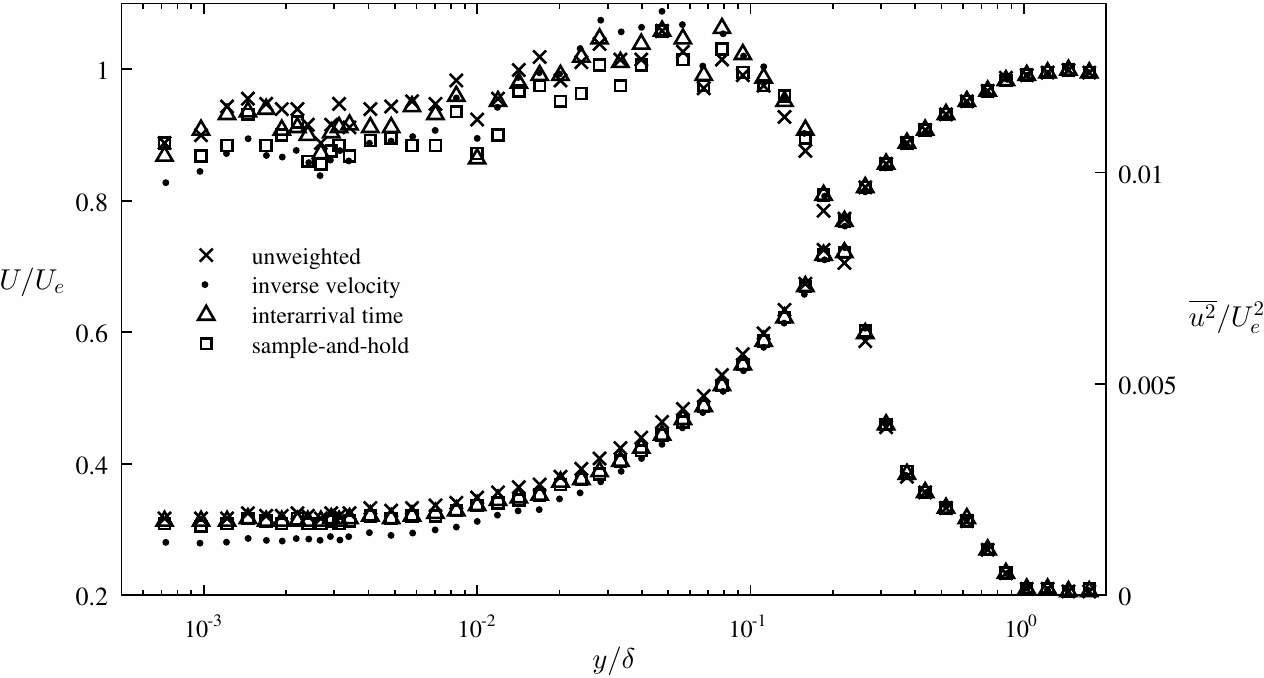}
	\caption{Mean velocity $U$ and streamwise turbulence intensity $\overline{u^2}$ profiles corrected by inverse velocity magnitude and inter-arrival time weighting.  The wall-normal coordinate is normalized using the local 99\% boundary layer thickness, $\delta$.}
	\label{fig:weights}
\end{figure}

LDV measurements suffer from two significant distortions: velocity gradients across the measurement volume and velocity biasing \citep{durst1976principles, degraaff2001high}. The former is significant in regions where large velocity gradients are present across the measurement volume, as is the case near solid walls. The latter occurs because, in turbulent flows, more high velocity particles move through the measurement volume in a given period compared to low-velocity particles (assuming uniform seeding density). To correct for this bias, an inverse velocity weighting factor was used to correct mean statistics \citep{mclaughlin1973biasing}.  The mean streamwise velocity was estimated using the following relationship: 
\begin{equation}\label{eq:weight-U}
U = \frac{\sum\limits_{i=1}^{N}b_i u_i}{\sum\limits_{i=1}^{N} b_i},
\end{equation}
where $u_i$ is an individual velocity sample, $N$ is the total number of samples, and $b_i = 1/|u_i|$ is the weighting factor.  Similarly, the weighted streamwise turbulence intensity was estimated as:
\begin{equation}\label{eq:weight-u}
\overline{u^2}=\frac{\sum\limits_{i=1}^{N} b_i \left[u_i-U\right]^2}{\sum\limits_{i=1}^{N}b_i}.
\end{equation}
Alternative methods of correcting for velocity biasing are discussed and evaluated in \cite{herrin1993investigation}. These methods include using the inter-arrival time ($b_i = t_i-t_{i-1}$) as the weighting factor, or the sample-and-hold technique ($b_i=t_{i+1}-t_i$).  Figure \ref{fig:weights} shows how these different correction techniques affect a representative mean velocity ($U$) and streamwise turbulent intensity ($\overline{u^2}$) profile.  Inverse velocity weighting leads to the largest correction relative to the raw data in the near-wall region where the sampling is more intermittent (n.b. this is also consistent with previous studies).  However, the overall trends remain very similar in all cases and the correction is minimal in the outer region of the flow (see e.g. $y/\delta \ge 0.1$ in figure \ref{fig:weights}).

\begin{figure}
	\centering\includegraphics[scale=0.8]{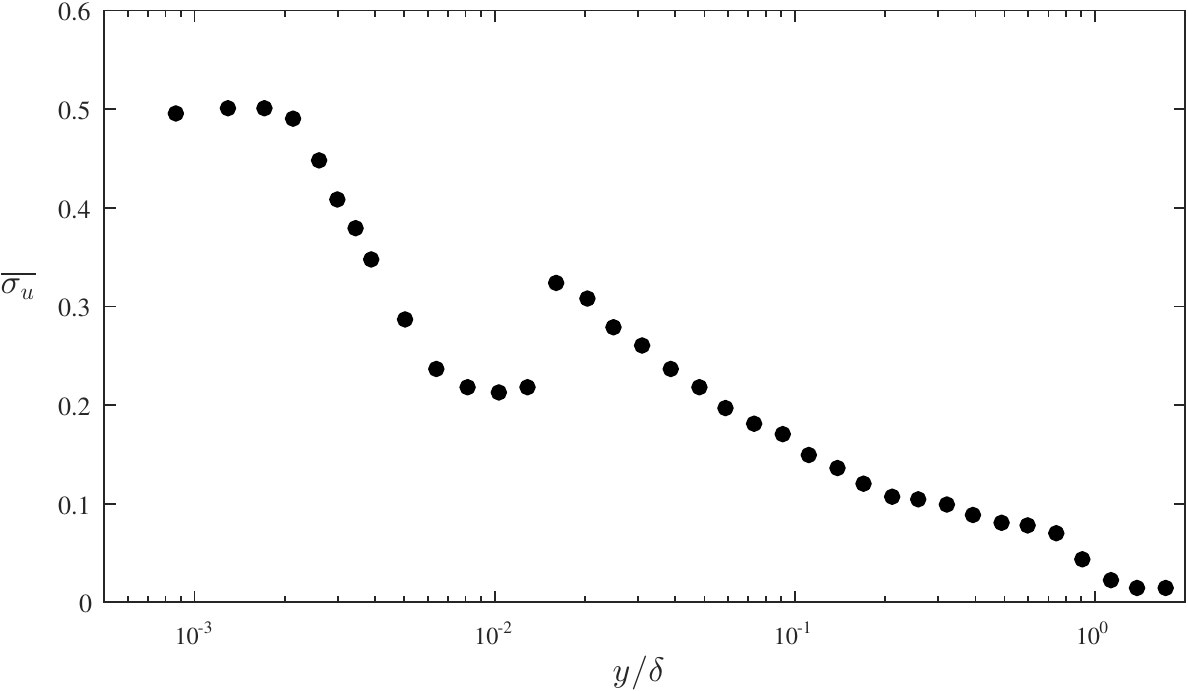}
	\caption{Profile showing standard error of mean velocity (in $\%$) over upstream smooth wall section. The discontinuity in $\overline{\sigma_u}$ between $y/\delta = 0.01$ and $0.02$ corresponds to the location where the measurement duration for each point was reduced from 45 minutes to 10 minutes.}
	\label{fig:uncertainty}
\end{figure}

While the corrections accounting for velocity biasing do introduce uncertainty, this uncertainty is likely to be correlated across all cases.  Therefore, for comparison across cases, the uncertainty in mean velocity is taken to be the larger of the instrument uncertainty ($0.1$\%, MSE Inc.) and the standard error, given by $\overline{\sigma_u}= \sigma_u/\sqrt{N}$, where $\sigma_u$ is the standard deviation of the measurements.  In all cases, the standard error was the larger contributer to uncertainty in the near-wall region due to the limited number of samples acquired.  As shown in figure \ref{fig:uncertainty}, the standard error was typically $\overline{\sigma_u} \le 1$\% across the entire boundary layer.

The friction velocity, $u_\tau$, over the smooth-wall section upstream of the porous cutout was estimated by fitting the following relationship to the near-wall mean velocity measurements: $U(y) = (u_\tau^2/\nu)(y^\prime + y_0)$, in which $y_0$ represents an offset from the nominal wall location where the data rate falls to zero, $y^\prime = 0$.  The smooth-wall velocity profiles reported below correct for this offset.  In other words, the true wall-normal distance is assumed to be $y = y^\prime + y_0$, such that the near-wall velocity profile is consistent with the theoretical relation $U^+ = y^+$ (see figure \ref{fig:dev-u}).  No such correction was made for the porous substrate.  For reference, the friction velocity upstream of the porous section was estimated to be $u_\tau = 2.3 \pm 0.05$ cm/s. The uncertainty is estimated by fitting the relationship in the viscous sublayer to different ranges of 5-10 points near the wall.  The offset was $y_0 = 40\mu$m, which translates into approximately one viscous length scale.  Note that the above estimate for the friction velocity is also consistent with estimates obtained from fits to the logarithmic region in the mean velocity profile using $\kappa =0.39$ \citep{marusic2013logarithmic}. The estimates for $u_\tau$ derived above also agreed with the third method utilizing the velocity gradient. See figure~\ref{fig:log-law} and related discussion in \S\ref{sec:log-shift} for further detail.

\subsection{Reynolds number ranges and spatio-temporal resolution}\label{sec:Re}
The 99\% boundary layer thickness, estimated via interpolation, was $\delta = 6.83 \pm 0.07$ cm for the smooth-wall profile, and so the friction Reynolds number was $\delta^+ = u_\tau \delta/\nu = 1690 \pm 70$ upstream of the porous section.  Over the porous sections, the boundary layer thickness increased; the maximum measured value was $\delta = 11.2 \pm 0.1$ cm.  As a result, the Reynolds number based on the nominal free-stream velocity ranged from $Re = U_e \delta/\nu = 42600-69900$.  The estimated friction velocity $u_\tau = 2.3 \pm 0.05$ cm/s translates into a viscous length-scale $\nu/u_\tau \approx 40 \mu$m and viscous time-scale $\nu/u_\tau^2 \approx 1.7$ ms.  Thus, the 16 $\mu$m precision of the traverse provides adequate vertical resolution for profiling purposes.  However, the LDV measurement volume ($150 \mu$m in $y$) extends across 3 viscous units in the wall-normal direction, which means that the near-wall measurements reported below suffer from distortion due to velocity gradients.  In the near-wall region, the LDV sampling frequency was approximately 0.5 Hz, which corresponds to an average sampling time of 1200 viscous units. In the outer region of the flow ($y/\delta \gtrsim 0.05$), the sampling frequencies were as high as 50 Hz, which translates into an average sampling time of 12 viscous units.  In other words, time-resolved velocity measurements are only expected in the outer region of the flow.

\subsection{Porous Substrates}\label{sec:foams}
\begin{figure}
\centering
\includegraphics[width=\textwidth]{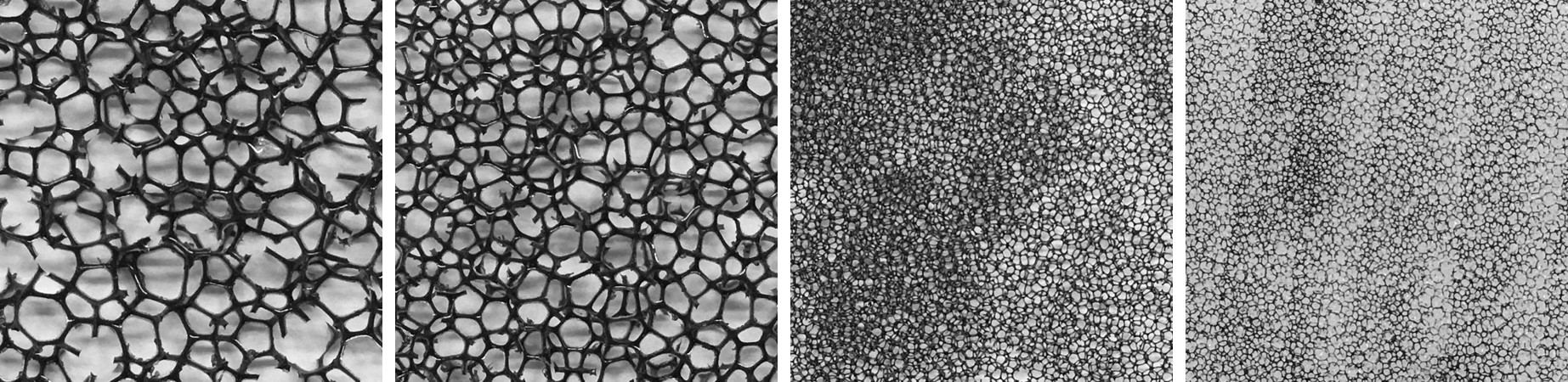}
\caption{Photographs showing thin sheets of the 10, 20, 60, and 100 ppi foams (from left to right).  Each image represents a 2 cm $\times$ 2 cm cross section.}
\label{fig:foam}
\end{figure}

Boundary layer measurements were made adjacent to four different types of open-cell reticulated polyurethane foams. Per the manufacturer, the porosity of all the foams was $\epsilon \approx 0.97$.  This was confirmed to within $0.5$\% via measurements that involved submerging the foams in water to measure solid volume displacements. The nominal pore sizes corresponded to 10, 20, 60, and 100 pores per inch (ppi, see figure \ref{fig:foam}). Pore size distributions for each foam were estimated from photographs of thin foam sheets via image analysis routines in Matlab (Mathworks Inc.).  The measured average pore sizes ranged from $s = 2.1 \pm 0.3$ mm for the 10 ppi foam to $s = 0.29 \pm 0.02$ mm for the 100 ppi foam (see Table~\ref{tab:foam}). These measurements are generally within $\pm 20\%$ of the nominal pore sizes.  \red{Pore size measurements for the 10 and 20 ppi foam were also made using precision calipers. These caliper-based measurements were consistent with the imaging-based estimates to within uncertainty ($s = 2.2 \pm 0.1$mm for 10 ppi and $s = 1.7 \pm 0.1$ mm for 20 ppi, where uncertainties correspond to standard error).
	
Note that all of the pore sizes discussed above and listed in Table.~\ref{tab:foam} correspond to the \textit{exposed} streamwise-spanwise plane of the foam.  Caliper-based measurements suggest that the pore structure may be anisotropic. For the 10 ppi foam, average pore sizes were approximately 14\% larger than the listed values in the spanwise-wall normal plane of the foam ($2.5 \pm 0.1$ mm) and 21\% larger in the streamwise-wall normal plane ($2.7 \pm 0.1$ mm). Similarly, for the 20 ppi foam, average pore sizes were approximately 10\% larger in the spanwise-wall normal plane and 23\% larger in the streamwise-wall normal plane.}  

Another important length scale arises from the permeability, $k$, of the porous medium.  Specifically, $\sqrt{k}$ determines the distance to which the shear penetrates into the porous medium \citep{battiato2012self}. \red{Permeabilities were estimated from pressure drop experiments using Darcy's law. These estimates ranged from $k = 6.6 \pm 0.6 \times 10^{-9}$ m$^2$ for the 100 ppi foam to $k = 46 \pm1 \times 10^{-9}$ for the 10 ppi foam.}

Based on the friction velocity measured upstream of the plate, the inner-normalized pore sizes range from $s^+ \approx 7$ for the 100 ppi foam to $s^+ \approx 52$ for the 10 ppi foam.  \red{Similarly, the Reynolds number based on permeability varies between $Re_k = \sqrt{k}u_\tau / \nu \approx 2.0 $ for the 100 ppi foam to $Re_k \approx 5.3 $ for the 10 ppi foam. The baseline thickness tested for all foams was $h=12.7$ mm.  For the 20 ppi foam, two additional thicknesses, $h = 6.35$ mm and $h = 25.4$ mm, were also considered. This means that the ratio of foam thickness to average pore size ranged between $h/s = 4.3$ for the thin 20 ppi foam to $h/s = 44$ for the 100 ppi foam. Finally, keep in mind that despite having the same nominal pore sizes, the 10 and 60 ppi foams tested here are not identical to those tested by \citet{manes2011turbulent}. For example, the 10 ppi foam tested by \citet{manes2011turbulent} had a pore size ($s=3.9$ mm) approximately twice that of the 10ppi foam used here, and a permeability ($k = 160 \times 10^-9$ m$^2$) almost four times higher.} Table~\ref{tab:foam} lists physical properties for all the foams tested in the experiments, along with related dimensionless parameters.

\begin{table}
  \begin{center}
  \begin{tabular}{lcccccccccc}
  Foam \hspace{1cm}	& $k$($10^{-9}$m$^2$)  & $\epsilon$ & $s$ (mm)  & $Re_k$ & $s^+$ &$h/\sqrt{k}$ &$h/s$ \\[0.1cm] 
  10 ppi 	& $46\pm1$   	& $0.976 \pm 0.003$ & $2.1 \pm 0.3$	& $5.3$ & 52	& 59 & 6	\\ 
  			& (160)  		& 					& (3.9) 		& 		&	 	&	 & 		\\[0.1cm] 
  20 ppi 	&				&  					& 				& 		&		& 73 & 8.5	\\ 
  20 ppi thin & $30\pm2$	& $0.972 \pm 0.003$ & $1.5 \pm 0.2$	& $4.3$ & 37	& 37 & 4.3  \\ 
  20 ppi thick &			&  					& 				&		&		&147 & 17 	\\[0.1cm] 
  60 ppi 	& $7.9\pm0.6$ 	& $0.965 \pm 0.005$ &$0.40 \pm 0.03$& $2.2$ & 10 	&143 & 32   \\ 
   	 		& (6) 			& 					& (0.5)			& 		& 		&	 &	 	\\[0.1cm]
  100 ppi 	& $6.6\pm 0.6$ 	& $0.967 \pm 0.005$ &$0.29 \pm 0.02$& $2.0$ & 7		&156 & 44	\\
  \end{tabular}
  \caption{\red{Permeability ($k$), porosity ($\epsilon$), average pore sizes ($s$), and related dimensionless parameters for tested foams.  Permeability and pore size values from \cite{manes2011turbulent} are noted in parenthesis for the 10 and 60 ppi foams. Note that $s^+ = s u_\tau /\nu$ and $Re_k = \sqrt{k} u_\tau/\nu$ are defined using the friction velocity upstream of the porous section. Porosity ($\epsilon$) was estimated from solid volume displacement in water and permeability was estimated from pressure drop experiments.}}
  \label{tab:foam}
  \end{center}
\end{table}

\section{Results}\label{sec:results}

\subsection{Boundary layer development}\label{sec:development}
First, we consider boundary layer development over the porous foams. The results presented below correspond to the 20 ppi foam \red{of thickness $h=12.7$mm}. Similar trends were observed for all the porous substrates.

\begin{figure}
	\centering
	\includegraphics[scale=0.8]{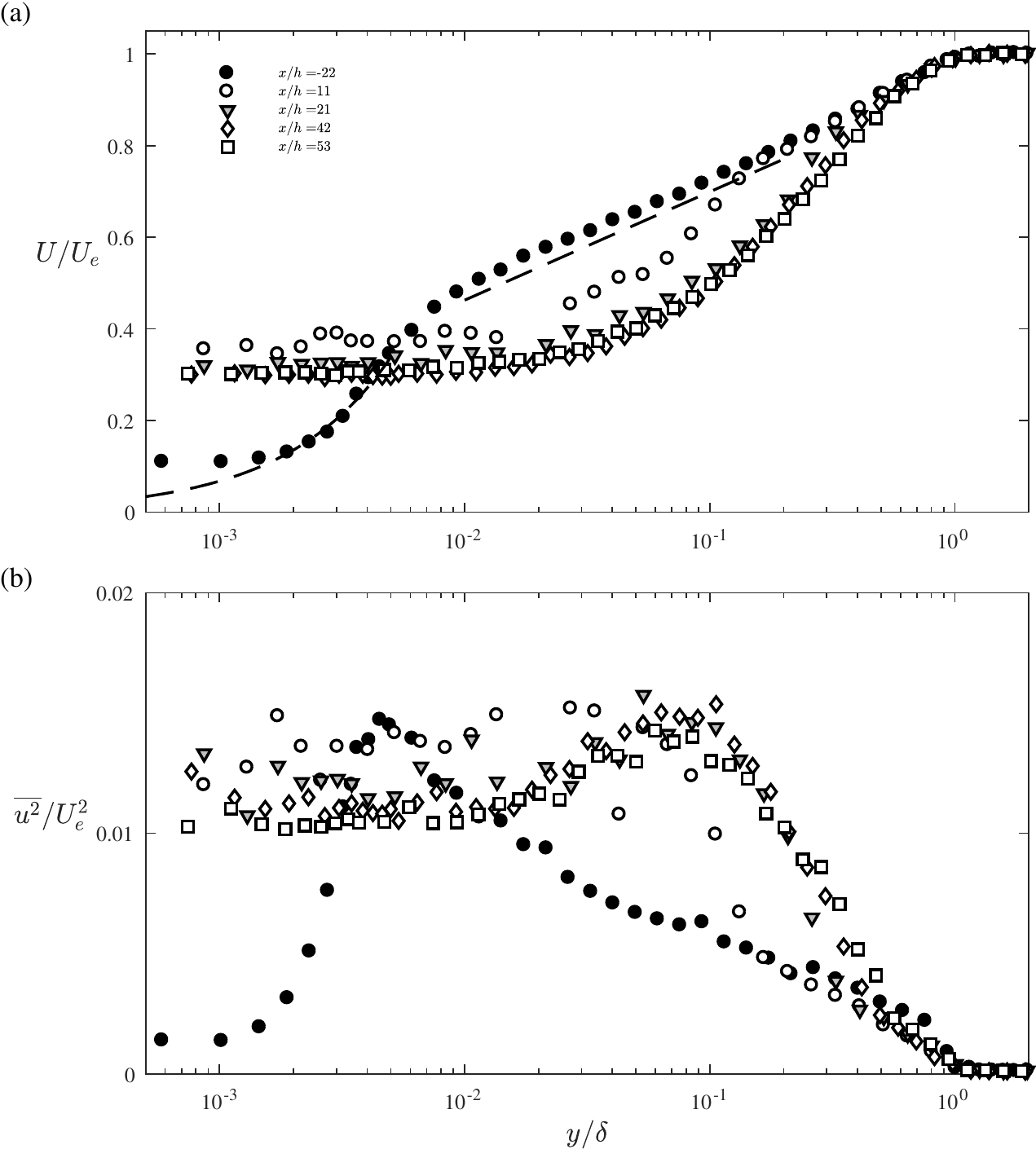}
	\caption{Streamwise mean velocity (a) and turbulence intensity (b) measured at streamwise locations $x/h={-21.5,11,21,42,53}$ relative to the transition from smooth wall to porous substrate.  These data correspond to the 20 ppi foam with $h=12.7$mm.  The dashed lines correspond to a linear profile of the form $U^+ = y^+$ in the near-wall region and a logarithmic profile of the form $U^+ = (1/\kappa)\ln(y^+)+B$ in the overlap region, with $\kappa = 0.39$ and $B = 4.3$. The friction velocity was estimated from fitting a linear slope to the near-wall measurements.}
	\label{fig:dev-u}
\end{figure}

As illustrated schematically in figure \ref{fig:diag}, the transition from the smooth wall to the porous substrate leads to the development of an internal layer, which starts at the transition point and grows until it spans the entire boundary layer thickness. When this internal layer reaches the edge of the boundary layer, a new equilibrium boundary layer profile is established.  This new profile reflects the effects of the porous substrate.  Internal layers have been studied extensively in the context of smooth to rough wall transitions in boundary layers \citep[e.g.,][]{antonia1971response,jacobi2011new}. In particular, previous literature suggests that turbulent boundary layers adjust relatively quickly for transitions from smooth to rough walls; the adjustment occurs over a streamwise distance of $O(10\delta)$.  The profiles of mean velocity ($U/U_e$) and streamwise intensity ($\overline{u^2}/U_e^2$) shown in figure \ref{fig:dev-u} suggest that the adjustment from smooth to porous wall velocity profiles occurs over a similarly short streamwise distance.

Upstream of the porous section, the measured profiles are consistent with previous smooth wall literature.  In the near-wall region ($y/\delta < 5\times 10^{-3}$), the measurements agree reasonably well with the fitted linear velocity profile $U^+ = y^+$.  The measured mean velocities are higher at the first two measurement locations, which can be attributed to the bias introduced by velocity gradients across the LDV measurement volume. In the overlap region ($0.02 \le y/\delta \le 0.2$), the mean velocity profile is consistent with the logarithmic law: $U^+ = (1/\kappa)\ln (y^+) + B$.  The streamwise intensity profile shows the presence of a distinct inner peak at $y/\delta \approx 0.006$, or $y^+ \approx 10$, which is also consistent with previous studies.

The presence of the porous substrate substantially modifies the mean velocity and streamwise intensity profiles.  Figure \ref{fig:dev-u} shows clear evidence of substantial slip at the porous interface ($U(y\approx 0) > 0.3U_e$).  Farther from the interface, there is a velocity deficit relative to the upstream, smooth wall profile. This velocity deficit increases with distance along the porous substrate, and appears to saturate for the final two profiles measured at $x/h \ge 42$. The mean velocity profiles collapse together for $y/\delta \ge 0.5$, suggesting that the outer part of the wake region remains unchanged over the porous substrate.

Consistent with the mean velocity profiles, the streamwise intensity profiles also show a substantial departure from the smooth case.  Although there is some scatter in the measurements closest to the interface, the inner peak is replaced by an elevated plateau at $\overline{u^2}/U_e^2 \approx 0.01$, which extends from the porous interface to $y/\delta \approx 0.01$.  Further, an outer peak appears near $y/\delta \approx 0.1$. The origin of this outer peak is discussed further in \S\ref{sec:poresize} below. In general, the streamwise intensity profiles also converge for $x/h \ge 42$.

\begin{figure}
	\centering\includegraphics[scale=0.8]{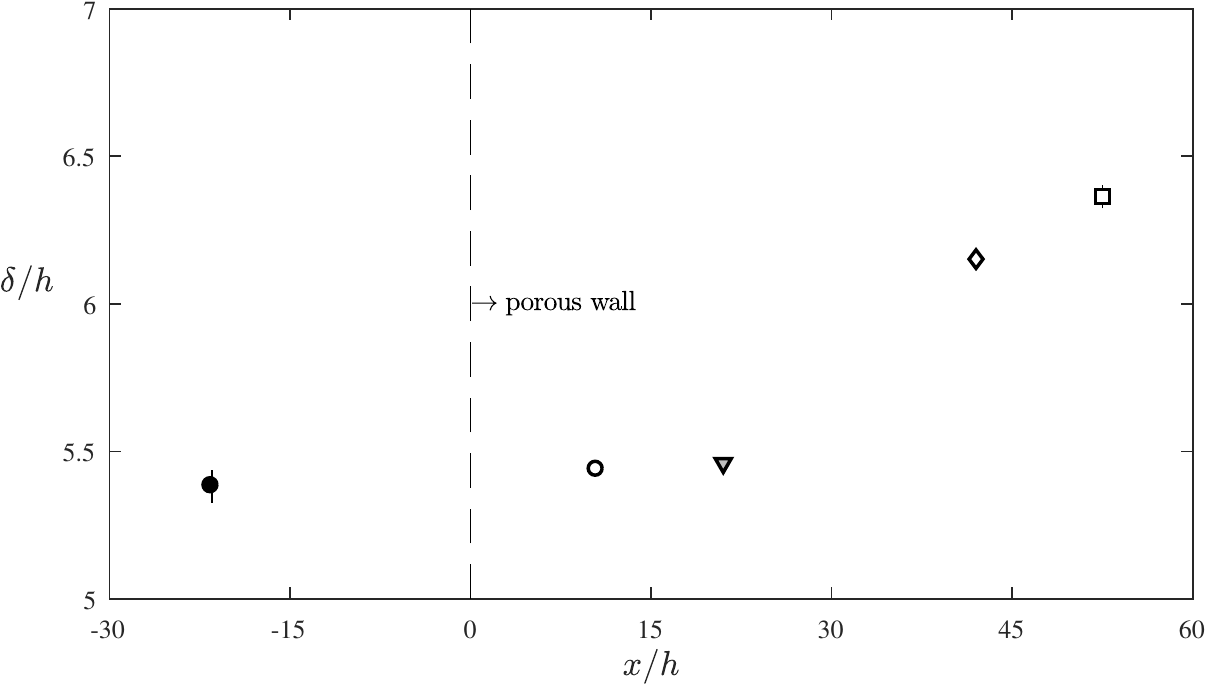}
	\caption{Normalized boundary layer thickness $\delta/h$ for the 20 ppi foam measured at streamwise locations $x/h={-21.5,11,21,42,53}$ relative to the substrate transition point.}
	\label{fig:dev-delta}
\end{figure}

The streamwise evolution of the boundary layer thickness is presented in figure \ref{fig:dev-delta}.  Boundary layer growth over the first two measurement locations ($x/h \le 21$) past the smooth to porous transition is relatively slow and appears unchanged from the smooth wall boundary layer, suggesting that the internal layer does not yet span the boundary layer thickness at these locations.  For the last two measurement locations, $x/h \geq 42$, the boundary layer thickness grows much faster, suggesting that the flow adjustment is complete and that the effects of the porous substrate extend across the entire boundary layer.  These observations are consistent with the mean velocity and streamwise intensity profiles shown in figure \ref{fig:dev-u}.  As an example, the profiles at measurement location $x/h = 11$ show that the internal layer only extends to $y/\delta \approx 0.2$.  For $y/\delta \ge 0.2$, the mean velocity and streamwise intensity collapse onto the smooth-wall profiles.

Development data for the remaining foams with are not presented here for brevity. \red{For all the foams of thickness $h \le 12.7$ mm}, the velocity measurements and boundary layer thickness data suggest that flow adjustment is complete by the measurement station at $x/h = 42$. Since the incoming boundary layer thickness is $\delta \approx 5.5 h$, the streamwise adjustment happens over $x \approx 8\delta$, which is consistent with previous literature on the transition from smooth to rough walls. \citep{antonia1971response} \red{However, this analogy to flow adjustment over rough-wall flows breaks down for the the thick 20 ppi foam with $h = 25.4$ mm. For the thick foam, flow adjustment was not complete even at the last measurement location (see results presented in \S \ref{sec:thickness}). This observation is still in broad agreement with the results presented above since the last measurement location only yields a dimensionless development length of $x/h \approx 26$ for the thick foam, while figure \ref{fig:dev-delta} suggests that $x/h \ge 30$ is required for adjustment.}

\red{The specific setup considered here may also be seen as flow over a backward facing step, with the region beyond the step filled with a highly porous ($\epsilon > 0.96$) and permeable ($Re_k>1$) material. Step Reynolds numbers in the present experiment range from $Re_h=U_e h/\nu = 4-16\times 10^3$. DNS by \citet{le1997direct} and LDV measurements by \citet{jovic1994backward} at $Re_h=5100$ for a canonical (i.e., unfilled) backward facing step indicate that the mean velocity profile behind the step does not return to a log-law for $x/h=20$ beyond the step. This is comparable to the development length observed here over the flush-mounted highly porous substrates ($x/h>30$).  Thus, it appears that both $\delta$ and $h$ play a role in dictating flow adjustment for the system considered here.} Unfortunately, since the present study was limited to four measurement locations along the porous substrate, there is insufficient spatial resolution in the streamwise direction to provide further insight into the scaling behaviour of boundary layer adjustment and growth over porous substrates.

\subsection{Effect of pore size}\label{sec:poresize}

\begin{figure}
	\centering
	\includegraphics[scale=0.8]{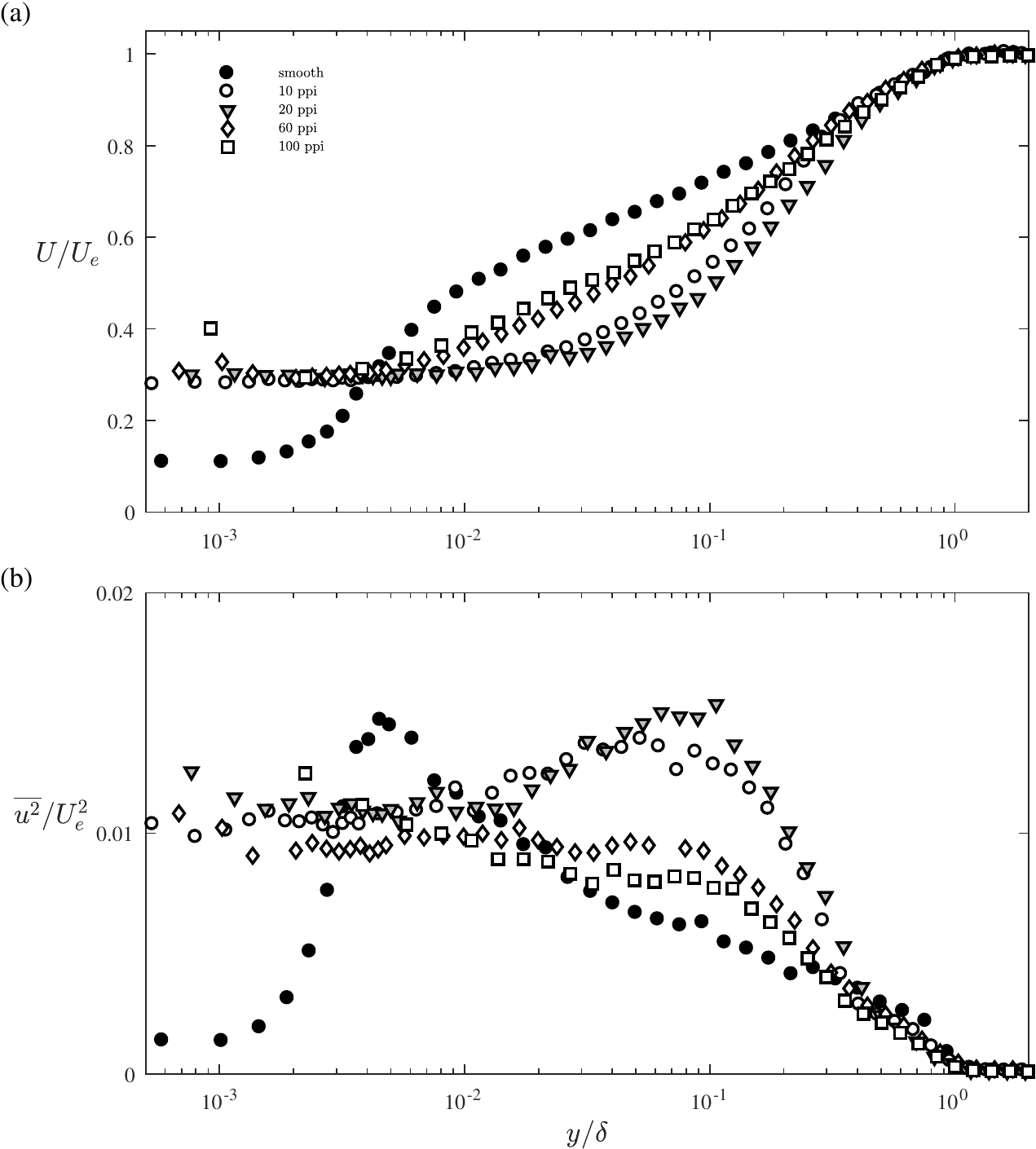}
	\caption{Mean velocity (a) and turbulence intensity (b) profiles for the smooth wall and for all the porous foams at $x/h=42$.}
	\label{fig:ppi-u}
\end{figure}

Next, we consider the effect of varying pore size on the fully-developed boundary layer profiles measured at $x/h = 42$ \red{for the foams of thickness $h = 12.7$ mm}. Figure \ref{fig:ppi-u} shows the measured mean velocity and streamwise intensity for each of the foams tested, together with the smooth-wall profile measured upstream of the porous section.

The mean velocity profile over the porous foams is modified in two significant ways with respect to the smooth wall profile taken upstream. First, there is a substantial slip velocity near the porous substrate. This slip velocity is approximately 30$\%$ of the external velocity across all substrates, with little dependence on pore size.  The observed slip velocity is consistent with the DNS results of \citet{breugem2006influence}, who observed a slip velocity of approximately $30\%$ for their highest porosity case ($\epsilon=0.95$, $Re_k=9.35$). Note that the properties for the foams tested here (listed in table \ref{tab:foam}) are similar to the properties of the porous substrate considered in the simulations.  Second, there is a mean velocity deficit relative to the smooth-wall case from $0.004\leq y/\delta \leq 0.4$.  This deficit generally increases with increasing pore size, though there is some non-monotonic behaviour. The maximum deficit relative to the smooth-wall profile is approximately 15$\%$ for the 100 ppi foam ($s^+ = 7$) and this increases to almost 50$\%$ for the 20 ppi foam ($s^+ = 37$).  However, the deficit for the foam with the largest pore sizes (10 ppi, $s^+ = 52$) is smaller than the deficit over the 20 ppi foam. In the outer wake region ($y/\delta > 0.5$), all mean velocity profiles collapse onto the canonical smooth-wall profile again.

The streamwise turbulence intensity profiles, plotted in figure \ref{fig:ppi-u}b, show that the inner peak disappears for all the porous foams. Instead, there is a region of elevated but roughly constant intensity that extends from the porous interface to $y/\delta \approx 0.01$.  For $y/\delta > 0.01$, the intensity profiles show a strong dependence on pore size.  For the smallest pore sizes, the streamwise intensity either decreases slightly (100 ppi) or stays approximately constant until $y/\delta \approx 0.1$ (60 ppi).  For foams with larger pore sizes, the intensity increases and a distinct outer peak appears near $y/\delta \approx 0.1$. All the profiles collapse towards smooth-wall values for $y/\delta \ge 0.5$.  Consistent with the mean velocity measurements, the streamwise intensity profiles also show non-monotonic behaviour with pore size.  While the streamwise intensity in the outer region of the flow generally increases with pore size, the magnitude of the outer peak is higher for the 20 ppi foam compared to the 10 ppi foam.  Velocity spectra, presented in \S\ref{sec:spectra} below, provide further insight into the origin of this outer peak.

\begin{figure}
	\centering
	\includegraphics[scale=0.8]{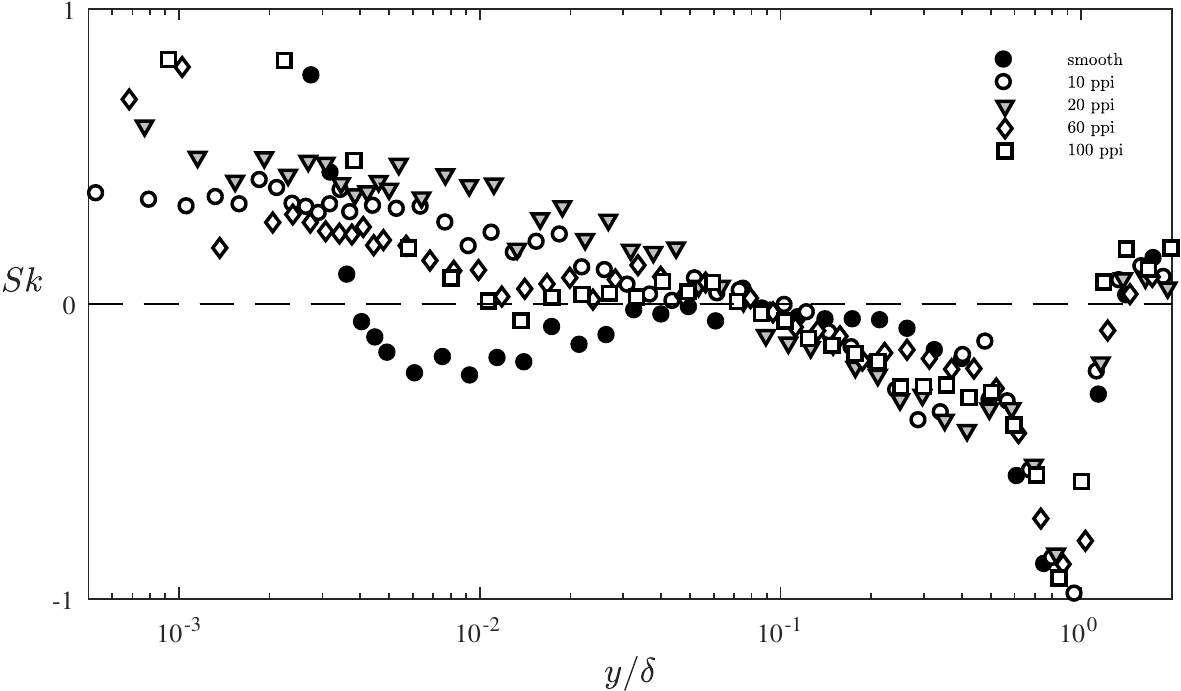}
	\caption{Wall-normal profiles of skewness ($Sk$) for the smooth wall and for all the porous foams at $x/h = 42$.}
	\label{fig:skew}
\end{figure}

Figure \ref{fig:skew} shows skewness profiles over the smooth wall and porous foams.  Consistent with previous measurements at comparable Reynolds number \citep[e.g.,][]{mathis2011relationship}, the skewness over the smooth wall is negative or close to zero across much of the boundary layer ($0.004 < y/\delta < 1$).  In contrast, over the foam substrates, the sign of the skewness is positive until $y/\delta \approx 0.1$. Interestingly, the location of this change in sign for the skewness corresponds to the location of the outer peak in streamwise intensity profiles.  Further, the magnitude of the skewness generally increases with pore size, which is consistent with the intensity measurements (the 10 and 20 ppi cases again show non-monotonic behaviour).  These observations are particularly interesting given the intrinsic link between skewness and amplitude modulation \citep{schlatter2010quantifying,mathis2011relationship}, and suggest that structures responsible for the outer peak in streamwise intensity over the foams may have a modulating effect on the interfacial turbulence. This possibility is discussed in greater detail in \S\ref{sec:amplitude-mod} below.

\subsection{Effect of substrate thickness}\label{sec:thickness}

\begin{figure}
	\centering
	\includegraphics[scale=0.8]{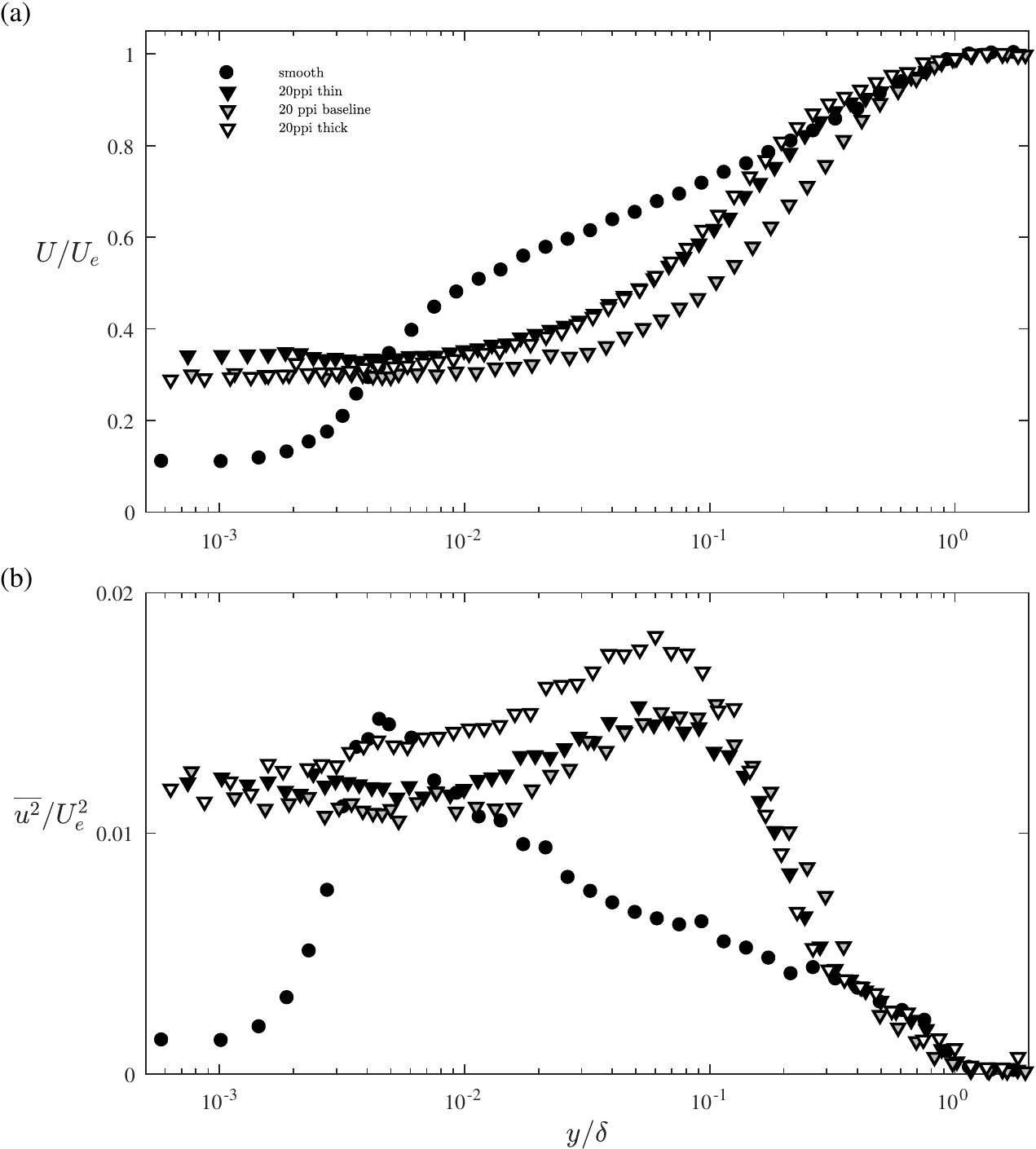}
	\caption{Mean velocity (a) and turbulence intensity (b) profiles for the smooth wall and for the 20ppi foam of varying thickness at the same physical location.}
	\label{fig:thick}
\end{figure}

\red{The thickness of the porous substrate, $h$, is another important parameter that dictates flow behaviour. Figure \ref{fig:thick} shows the measured mean velocity and streamwise intensity profiles over 20 ppi foams of varying thickness, $h = 6.35, 12.7$ and $25.4$ mm. In dimensionless terms, these thicknesses correspond to $h/s = 4.3, 8.5$ and $17$, respectively, where $s = 1.5 \pm 0.2$ mm is the average pore size (Table~\ref{tab:foam}).  Note that the measurements shown in figure~\ref{fig:thick} are for the same physical location, $x = 53.3$cm, which yields normalized distances increasing from $x/h = 21$ for the thickest foam to $x/h = 84$ for the thinnest foam.  The foam with thickness $h=12.7$mm is considered the baseline case, since velocity measurements made over this foam have already been discussed in \S\ref{sec:development} and \S\ref{sec:poresize}. The foams of thickness $h = 6.35$mm and $h = 25.4$mm are referred to as the thin and the thick foam, respectively. As noted in \S\ref{sec:development}, the flow was fully developed for the thin and baseline foams, but still developing for the thick foam.

Mean velocity profiles for both the thick and thin foam show similar slip velocities at the interface compared to the baseline foam of thickness $h=12.7$ mm ($\approx 0.3 U_e$), though the velocity measured over the thin foam is slightly higher ($\approx 0.35U_e$).  In general, the mean profile for the thin foam is consistently higher than that for the baseline foam, and collapses onto the smooth-wall profile above $y/\delta > 0.2$ (black triangles in figure~\ref{fig:thick}a). In contrast, the mean profile for the thicker foam (white triangles in figure~\ref{fig:thick}a) is closer to the baseline case in the near-wall region $y/\delta < 0.005$. However, a little farther from the wall, the mean profile for the thick foam diverges from that for the baseline foam.  The thick foam mean profile shows a smaller velocity deficit in the region $y/\delta \approx 0.01$ to $y/\delta \approx 0.1$ compared to the baseline case, and verges on the smooth-wall profile for $y/\delta \ge 0.2$.

The streamwise turbulence intensity profiles plotted in figure~\ref{fig:thick}b show that $\overline{u^2}/U_e^2$ is similar near the interface for all three foams. In fact, the baseline and thin foams show very similar intensity profiles through most of the boundary layer, barring two minor differences. First, the outer peak in streamwise intensity appears closer to the interface for the thin foam. Second, the thin foam profile shows a better collapse onto the smooth-wall profile for $y/\delta \ge 0.3$. The turbulence intensity profile for the thick foam also collapses onto the smooth-wall profile for $y/\delta \ge 0.3$.  However, closer to the wall, $\overline{u^2}$ is much higher over the thick foam, and the outer peak in intensity also appears to move slightly closer to the interface.  These features are also seen in the velocity spectra presented below.  

In summary, for the thin foam, the mean velocity and streamwise intensity measurements both collapse onto the smooth-wall profile for $y/\delta \ge 0.3$.  This suggests that the outer-layer similarity hypothesis proposed by Townsend for rough-walled flows \citep[see e.g.,][]{schultz2007rough} also holds for thin porous media. The mean profile for the thick foam is similar to that for the baseline foam close to the interface, but moves closer to the thin foam profile further from the interface.  However, the streamwise intensities are significantly higher for the thick foam compared to the thin foam for $y/\delta \le 0.1$, even in regions where the mean profiles show agreement. These discrepancies between the mean velocity and streamwise intensity profiles for the thick foam are consistent with a developing flow.}

\subsection{Velocity spectra}\label{sec:spectra}
\begin{figure}
	\centering
	\includegraphics[scale=0.85]{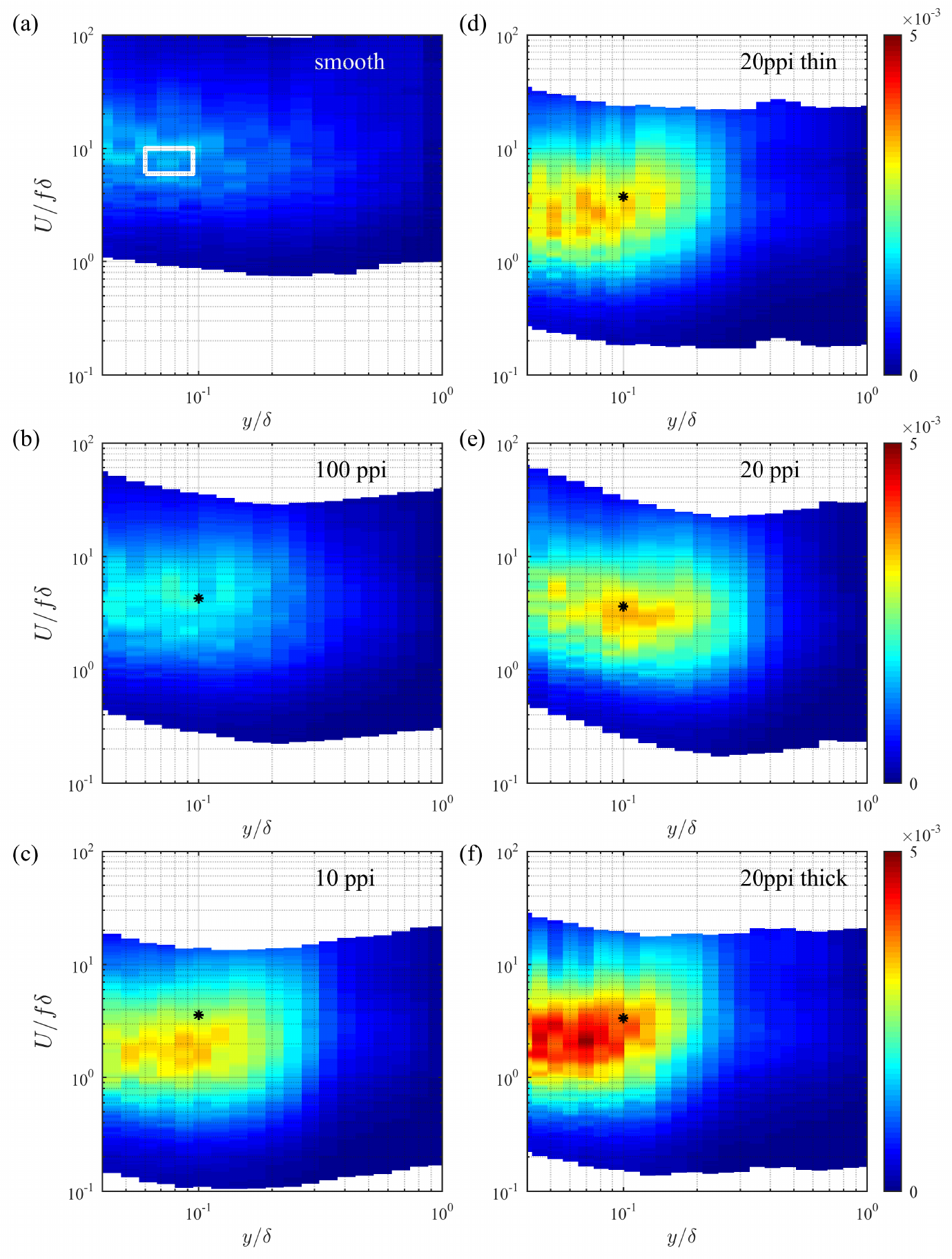}
	\caption{Contour maps showing variation in premultiplied frequency spectra \red{(normalized by $U_e^2$)} for streamwise velocity as a function of wall-normal distance $y/\delta$ over the smooth wall (a), the 100 ppi foam (b), the 10 ppi foam (c), the thin 20 ppi foam (d), the baseline 20 ppi foam (e), and the thick 20 ppi foam (f).  The spectra are plotted against a normalized streamwise length scale, $U/f\delta$, computed using Taylor's hypothesis.  The white box in (a) denotes the region typically associated with VLSMs while the markers ($*$) represent the nominal frequency $f_{KH}$ for structures resembling Kelvin-Helmholtz vortices. \red{The spectra  refer to the same physical measurement location, corresponding to $x/h=42$ for the foams with $h = 12.7$mm and $x/h=21$ and $84$ for the thick and thin 20 ppi foams, respectively.}}
	\label{fig:spectra}
\end{figure}

The premultiplied velocity spectra shown in figure \ref{fig:spectra} provide further insight into the origin of the outer peak in streamwise intensity observed over the porous substrates.  \red{As is customary in the boundary layer literature, the premultiplied spectrum is defined as $f E_{uu}$, where $f$ is the frequency and $E_{uu}$ is the power spectral density, normalized by $U_e^2$. This quantity is computed for each wall-normal location, and the results are compiled into the contour plots shown in figure \ref{fig:spectra}.} Due to the low data rates obtained near the interface, spectra are only shown for $y/\delta \ge 0.04$.  Note that the spectra are expressed in terms of a normalized streamwise wavelength estimated using Taylor's hypothesis, $U/(f\delta)$.

For the smooth wall case, there is evidence of weak very-large-scale motions (VLSMs), which is similar to results obtained in previous studies at comparable Reynolds number \citep{hutchins2007evidence}.  The box in the top left panel encompasses the spectral region typically associated with VLSMs, i.e. structures of length $6\delta - 10\delta$ ($U/f\delta = 6-10$) located between $y/\delta = 0.06$ and $y^+ = 3.9 \sqrt{Re_\tau}$ \citep[see][]{hutchins2007evidence,marusic2010predictive,smits2011high}. This box coincides with a region of elevated spectral density for the measurements.

Spectra for the porous substrates are different in several ways. For all the foams, the spectra are elevated over the frequency range that corresponds to structures with streamwise length scale $1\delta-5\delta$. The spectra are most energetic \red{at, or below,} wall-normal location $y/\delta \approx 0.1$ and remain elevated until $y/\delta \approx 0.3$. There is a marked increase in the spectral energy density from the 100 ppi foam to the 20 ppi foam, and little difference between the spectra for the 20 ppi and 10 ppi foams. Together, these features suggest that the outer peak in streamwise intensity observed over the porous foams \red{in figure~\ref{fig:ppi-u}b} is associated with large-scale structures of length $1\delta-5\delta$ that are distinct from VLSMs.

\red{Spectra for the 20ppi foam of different thickness, plotted in figure \ref{fig:spectra}e-f, show a substantial increase in energy for the thickest foam, which is consistent with the elevated streamwise intensity profiles shown in figure~\ref{fig:thick}b. For the thick foam, the energy is also concentrated closer to the interface compared to the thin and baseline foams with identical pore sizes.}

Note that the spectral features described above are consistent with previous experiments and simulations. \citet{manes2011turbulent} showed that the premultiplied frequency spectra for streamwise velocity peak at wavenumber $k_x\delta \approx 2-4$ over porous substrates, i.e. corresponding to structures of length $1.5\delta-3\delta$.  Similarly, the DNS carried out by \citet{breugem2006influence} indicated the presence of spanwise rollers with streamwise length-scale comparable to the total channel height.  Since these structures have been linked to a Kelvin-Helmholtz shear instability arising from the inflection point in the mean profile, it is instructive to consider the characteristic frequency associated with this mechanism. Drawing an analogy to mixing layers, \citet{white2007shear} and \citet{manes2011turbulent} suggested that the characteristic frequency for the Kelvin-Helmholtz instability can be estimated as $f_{KH} = 0.032 \overline{U}/\theta$ \citep{ho1984perturbed}, in which $\overline{U} = (U_p + U_e)/2$ is the \textit{average} velocity in the shear layer with $U_p$ being the velocity deep within the porous medium, and $\theta$ is the momentum thickness.  Estimates for this characteristic frequency, converted to streamwise length-scale based on the mean velocity at $y/\delta = 0.1$, $\lambda_{KH}/\delta = U(0.1)/f_{KH}\delta$, are shown in figure \ref{fig:spectra}.  These estimates assume $U_p \approx 0$, so that $\overline{U} \approx 0.5U_e$, and that the momentum thickness $\theta$ can be approximated based on the measured velocity profile in the fluid domain, i.e. for $y \ge 0$.  With these assumptions, the predicted length scales associated with the instability range from $\lambda_{KH}/\delta \approx 4.3$ for the 100 ppi foam to $\lambda_{KH}/\delta \approx 3.6$ for the 10 ppi foam, which is in reasonable agreement with the location of the peaks in the spectra. 

\red{There is an evident overprediction of length scale for both the 10 ppi foam (figure~\ref{fig:spectra}c) and the thick 20 ppi foam (figure~\ref{fig:spectra}c). This overprediction can be attributed to greater flow penetration into the porous medium for thicker foams with larger pore sizes. Greater flow penetration would create higher velocities inside the porous medium, $U_p$. This would increase $\overline{U}$ and $f_{KH}$, resulting in lower $\lambda_{KH}$. Greater flow penetration into the porous medium may also result in the inflection point moving closer to the interface. In this scenario, conversion from $f_{KH}$ to $\lambda_{KH}$ should be based on a lower mean velocity from $y/\delta < 0.1$. This would also result in lower $\lambda_{KH}$ compared to the estimates shown in figure~\ref{fig:spectra}.

To test whether the observed peaks in streamwise intensity and velocity spectra tracked the location of inflection points in the mean profile, $y_i$, finite-difference approximations of the second derivative of mean velocity, $(d^2U/dy^2)_{y_i} = 0$, were considered. This yielded inflection point locations ranging from $y/\delta \approx 0.03$ to $y/\delta \approx 0.12$ for the 10 and 20 ppi foams, which is broadly consistent with the location of energetic peaks in figure~\ref{fig:spectra}. However, the second derivatives estimated from experimental data were very noisy, and the inflection point locations were highly sensitive to the accuracy (i.e., order) of the finite-difference approximation and the lower limit in $y$ used for the estimates. As a result, exact inflection point locations are not presented here.}


\section{Discussion}\label{sec:discussion}
\subsection{Amplitude modulation}\label{sec:amplitude-mod}
As noted earlier, the skewness profiles shown in figure~\ref{fig:skew} strongly suggest that the amplitude modulation phenomenon that has received significant attention in recent smooth and rough wall literature \citep[e.g.,][]{mathis2009large,marusic2010predictive,mathis2013estimating,pathikonda2017inner} may also be prevalent in turbulent flows over permeable walls.

For smooth wall flows at high Reynolds number, it has been shown that the VLSMs prevalent in the logarithmic region of the flow can have a modulating effect on the small-scale fluctuations found in the near-wall region.  Specifically, it has been suggested that the turbulent velocity field in the near-wall region can be decomposed into a large-scale (or low-frequency) component $u_L$ that represents the near-wall signature of the VLSMs, and a `universal' (i.e. Reynolds number independent) small-scale component $u_S$ that is associated with the local near-wall turbulence. \red{Analysis of the resulting signals indicates that the filtered envelope of the small-scale activity, $E_L(u_S)$, obtained via a Hilbert transform \citep[for details, see][]{mathis2011relationship}, is strongly correlated with the large-scale signal. In other words, the single-point correlation coefficient:
\begin{equation}\label{eq:R}
R = \frac{\overline{u_L E_L(u_S)}}{\sqrt{\overline{u_L^2}}\sqrt{\overline{E_L(u_S)^2}}}
\end{equation}
tends to be positive in the near-wall region, which suggests that the small-scale signal $u_S$ is modulated by the large-scale signal $u_L$.

Since the local large-scale signal $u_L$ arises from VLSM-type structures centered further from the wall, }these observations have given rise to predictive models of the form
\begin{equation}\label{eq:modulation}
u = u^*(1+\beta u_{OL}) + \alpha u_{OL},
\end{equation}
where $u(y)$ is the predicted velocity at a specified location in the near-wall region, $u^*(y)$ is a statistically universal small-scale signal at that wall-normal location, $u_{OL}$ is the large-scale velocity measured in the outer region of the flow, and $\alpha(y)$ and $\beta(y)$ are superposition and modulation coefficients, respectively \citep{marusic2010predictive}.  Assuming that the universal small-scale signal can be obtained via detailed experiments or simulations carried out at low Reynolds numbers, equation (\ref{eq:modulation}) allows for near-wall predictions at much higher Reynolds numbers based only on measurements of $u_{OL}$ in the outer region of the flow. \red{Note that the modulation coefficient $\beta$ is similar to the single-point correlation coefficient $R$, but also accounts for changes in phase and amplitude of the large-scale signal from the measurement location to the near-wall region.  In other words, $\beta$ also accounts for the relationship between $u_{OL}$ and $u_L(y)$, which is thought to be Reynolds-number independent \citep{marusic2010predictive}.}

\red{Subsequent studies have shown rigorously that the correlation coefficient $R$ in (\ref{eq:R}) is intrinsically linked to the skewness of the velocity \citep{schlatter2010quantifying,mathis2011relationship,duvvuri2015triadic}, whereby positive correlations between scales ($R > 0$) translate into increased skewness. Thus, the increase in skewness observed in figure~\ref{fig:skew} near the porous interface suggests that $R>0$ in this region.}

Near-wall measurements made in the present study do not have sufficient time resolution to allow for a quantitative evaluation of the amplitude modulation phenomenon over porous substrates (i.e., a decomposition into small- and large-scale components). However, the increase in the skewness in the near-wall region over the porous substrates suggests that the large-scale structures responsible for the outer peak in streamwise intensity at $y/\delta \approx 0.1$ may have a modulating effect on the turbulence near the interface. This is particularly interesting given that the large-scale structures found over porous substrates are distinct from the VLSMs found over smooth walls, and that the small-scale turbulence near the porous interface may also be modified from the near-wall cycle \citep{robinson1991coherent,schoppa2002coherent} found over smooth walls.

\subsection{Logarithmic region}\label{sec:log-shift}
\begin{figure}
	\centering
	\includegraphics[scale=0.8]{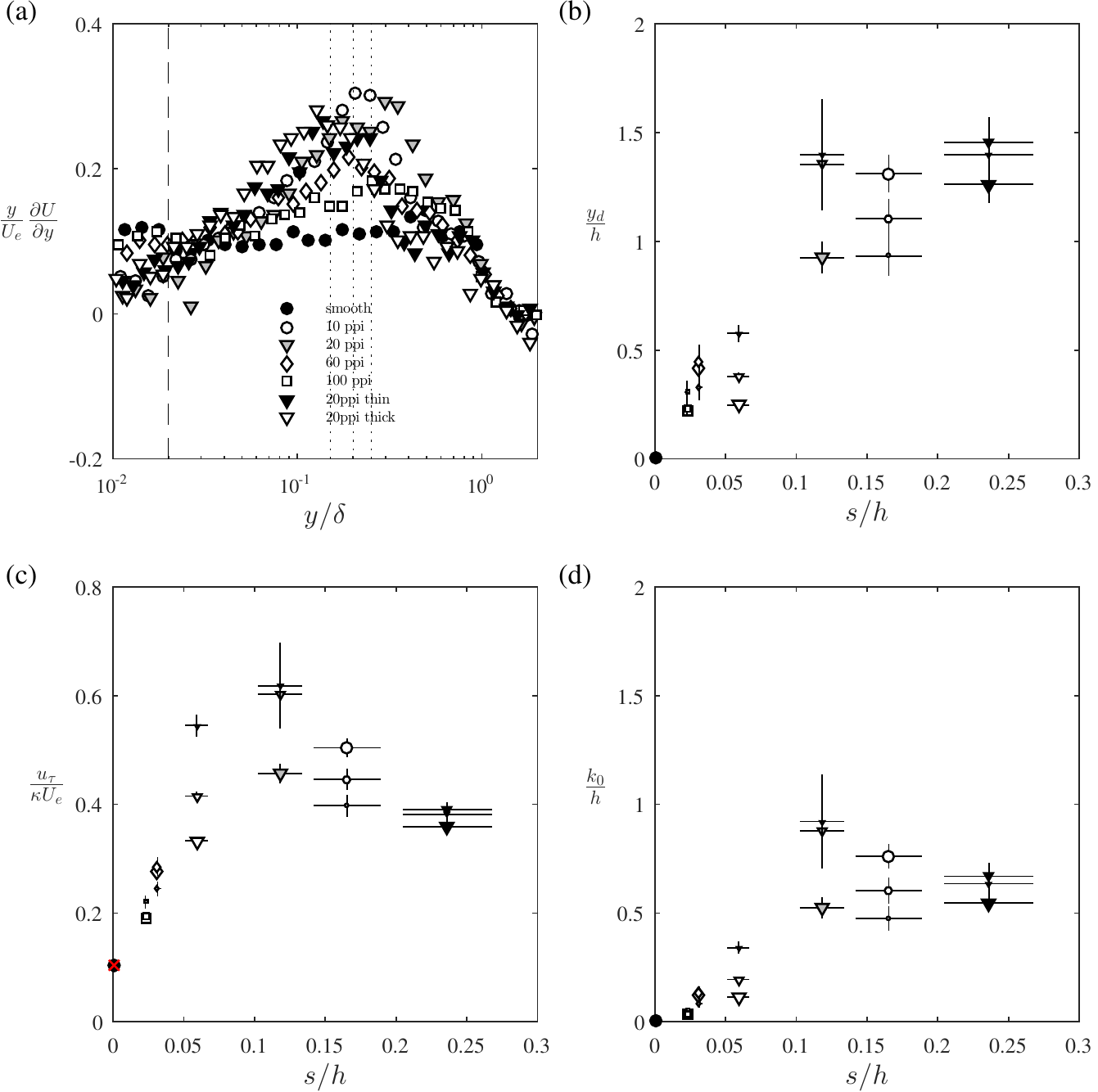}
	\caption{(a) Scaled velocity gradient $(y/U_e)\partial U/\partial y$ plotted as a function of $y/\delta$ for the smooth wall profile and porous foam data. Equation (\ref{eq:log-law-fit}) was fitted to these data to estimate the normalized displacement height, $y_d/h$, and the friction velocity weighted by the von Karman constant, $u_\tau/(\kappa U_e)$, shown in (b) and (c), respectively.  Using these estimates for $y_d$ and $u_\tau/\kappa$, the roughness height $k_0/h$, shown in (d), was evaluated from the velocity profiles using equation (\ref{eq:log-law}).  Dotted lines in (a) represent the upper limits of $y/\delta=0.16,0.20$ and $0.25$ employed in the fitting procedure. The dashed line represents the minimum lower limit, $y/\delta > 0.02$. Larger marker sizes in (b,c,d) denote higher values for the upper limit. The red cross in (c) represents the friction velocity estimated via a linear fit to the near-wall velocity measurements over the smooth wall. \red{Horizontal error bars in panels (b)-(d) represent uncertainty in pore size, $s$.}}
	\label{fig:log-law}
\end{figure}

Previous studies have devoted considerable effort to testing whether a modified logarithmic region of the form shown in equation (\ref{eq:log-law}) exists in turbulent flows over porous substrates.  While there is no definitive consensus on how the von Karman constant, $\kappa$, displacement height, $y_d$, and equivalent roughness height, $k_0$, vary with substrate properties, experiments and simulations generally show that the displacement and roughness heights increase with increasing permeability (or pore size).  \citet{manes2011turbulent} showed that empirical relationships of the form:
\begin{equation}\label{eq:ys}
y_d^+ = 15.1 Re_k - 13.5
\end{equation}
and
\begin{equation}\label{eq:k0}
k_0^+ = 6.28 Re_k - 9.82,
\end{equation}
where $Re_k = u_\tau \sqrt{k}/\nu$ is the permeability Reynolds number, led to reasonable fits for the experimental data obtained by \citet{suga2010effects} and \citet{manes2011turbulent}, but underestimated $y_d$ and $k_0$ for the simulations performed by \citet{breugem2006influence}.  The von Karman constant decreased relative to smooth wall values but demonstrated a complex dependence on both the permeability Reynolds number and the ratio of displacement height to boundary layer thickness, $y_d/\delta$ \citep{manes2011turbulent}.

Unfortunately, the present study did not involve independent measurements of the shear stress at the interface (or Reynolds' shear stress in the near-wall region) and so it is not possible to estimate the friction velocity and von Karman constant independently.  However, the velocity measurements can still be used to test whether a modified logarithmic region exists, and to estimate $u_\tau/\kappa$, $y_d$ and $k_0$.  Taking the partial derivative of equation (\ref{eq:log-law}) with respect to $y$ and rearranging yields:
\begin{equation}\label{eq:log-law-fit}
(y+y_d)\frac{\partial U}{\partial y} = \frac{u_\tau}{\kappa}.
\end{equation}
Following \citet{breugem2006influence} and \citet{suga2010effects}, we estimated $y_d$ as the value that forces $(y+y_d)\partial U/\partial y$ to be constant over specified ranges of $y/\delta$. Based on equation (\ref{eq:log-law-fit}), the resulting constant value for the weighted velocity gradient was assumed to be the friction velocity divided by von Karman constant, $u_\tau/\kappa$.  Using these estimates for displacement height and friction velocity, the roughness height was estimated directly from the velocity measurements using equation (\ref{eq:log-law}).

Since the fitting procedure described above relies on noisy velocity gradient data (see figure~\ref{fig:log-law}a), the uncertainty in the fitted values and sensitivity to fitting ranges was evaluated as follows.  First, the fitting procedure was carried out over the range $0.02 < y/\delta < 0.16$. The fit was then repeated with the lower limit sequentially increased by one to four measurement points.  The estimates of $y_d$, $u_\tau/\kappa$, and $k_0$ reported in figure~\ref{fig:log-law}(b-d) represent an average of the five different values obtained via this process and the error bars represent the standard error.  This entire procedure was then repeated for outer limits $y/\delta = 0.20$ and $y/\delta = 0.25$.  A similar process was used to evaluate $u_\tau/\kappa$ and $k_0$ for the smooth wall case with the displacement height constrained to be zero, $y_d = 0$. Estimates for $y_d$, $k_0$, and $u_\tau/\kappa$ are reported in table~\ref{tab:log-law}.

Note that the fitting procedure employed by \citet{manes2011turbulent} was also considered, where $y_d$ is estimated as the value that minimizes residuals between the measured velocity profile and equation (\ref{eq:log-law}) over specified ranges of $y/\delta$. The resulting fitted coefficients are then used to estimate $u_\tau/\kappa$ and $k_0$.  This process led to fitted values within the uncertainty ranges shown in figures~\ref{fig:log-law}b-d.

Assuming $\kappa = 0.39$, the fitting procedure described above led to a friction velocity estimate of $u_\tau = 2.31 \pm 0.02$ cm/s for the smooth wall profiles with outer limit $y/\delta = 0.15$. This is consistent with the value obtained via a linear fit to the near-wall mean velocity measurements, $u_\tau = 2.3 \pm 0.05$ cm/s.  The additive constant $B = -(1/\kappa)\ln k_0^+$ in equation (\ref{eq:log-law}) was estimated to be $B = 4.8 \pm 0.2$, which is slightly higher than the value, $B = 4.3$, reported in \citet{marusic2013logarithmic}, but still broadly consistent with previous literature.

As expected, the fitted log law constants for the porous substrates show a strong dependence on average pore size.  Figures~\ref{fig:log-law}(b-d) show that the displacement height, friction velocity, and roughness height generally increase with increasing pore size, though here is some evidence of saturation at the largest pore sizes.  Specifically, figure~\ref{fig:log-law}b suggests that the displacement height levels out above $y_d/h \approx 1$ for the \red{baseline 20 ppi foam, the 10 ppi foam, and the thin 20 ppi foam, for which $s/h > 0.1$}. This saturation in $y_d/h$ as a function of \red{normalized} pore size is accompanied by saturation, or perhaps slight decreases, in normalized friction velocity $u_\tau/\kappa U_e$ (figure~\ref{fig:log-law}c) and roughness height $k_0/h$ (figure~\ref{fig:log-law}d) \red{above $s/h > 0.1$}.

While the exact values of the log law constants shown in figure~\ref{fig:log-law} and table~\ref{tab:log-law} must be treated with some caution due to the uncertainty associated with the fitting procedure, the overall trends suggest the following physical interpretation.  The displacement height $y_d$ represents the level at which momentum is extracted within the porous medium \citep{jackson1981displacement}, or alternatively, the distance to which the turbulence penetrates into the medium \citep{luhar2008interaction} or the effective plane at which attached eddies are initiated \citep{poggi2004effect}.  For the least permeable substrates tested in the present study (i.e. the 100 ppi and 60 ppi foams), $y_d$ increases approximately linearly with average pore size. In other words, at this low permeability or thick substrate limit with $h/s \gg 1$, the distance to which turbulence penetrates into the porous medium increases with increasing pore size or permeability. Since the flow does not interact with the entire porous medium, the foam thickness $h$ does not play a role. However, with further increases in pore size \red{or decreases in porous medium thickness}, at some point the displacement height becomes comparable to the foam thickness $y_d \approx h$.  At this high permeability or thin substrate limit, turbulence penetrates the entire porous medium and the foam essentially acts as a roughness or obstruction. Results shown in figure~\ref{fig:log-law} suggest that \red{the baseline 20 ppi foam, the thin 20 ppi foam, and the 10 ppi foam, for which $h/s \le 10$}, may be approaching this thin substrate limit.

Figures~\ref{fig:log-law}b-d show that the normalized friction velocity and roughness height are strongly correlated with each other as well as the displacement height. Physically, the friction velocity is a measure of momentum transfer into the porous medium while the roughness height is a measure of momentum loss, or friction increase, due to the presence of the complex substrate.  As a result, correlation between $k_0$ and $u_\tau$ is unsurprising for boundary layer experiments carried out at constant free-stream velocity (n.b., for channel or pipe flow experiments, the friction velocity can be controlled independently by setting the pressure gradient).

The link between the displacement and roughness heights can be explained by considering the rough-wall literature.  For flows over conventional \textit{K}-type roughness, $k_0$ has been shown to depend on both the height of the roughness elements and the solidity $\lambda$, which is defined as the total projected frontal area per unit wall-parallel area \citep{jimenez2004turbulent}.  Similarly, for flows over porous media, $k_0$ can be expected to depend on the displacement height, which represents the thickness of porous medium that interacts with the flow, as well as the porous medium microstructure \citep[see also][]{jackson1981displacement,manes2011turbulent}.  In other words, a relationship of the form $k_0/y_d = f(\lambda)$ may be appropriate for turbulent flows over porous media.  Note that the physical link between the roughness and displacement heights is also evident in the empirical relationships for $y_d$ and $k_0$ shown in equations (\ref{eq:ys}-\ref{eq:k0}).

\begin{table}
	\begin{center}
		\begin{tabular}{lccccc}
		Substrate\hspace{1cm} & $u_\tau/\kappa$ (cm/s) & $y_d$ (mm) & $y_d/h$ & $k_0$ (mm) & $k_0/y_d$\\[0.1cm]
		smooth  & 5.8 	& 0    & 0 	  & $5.3\times 10^{-3}$ & $\infty$\\[0.1cm]
		10 ppi  & 27.5	& 14.2 & 1.1  & 7.8  & 0.55 \\[0.1cm]		
		20 ppi  & 32.9	& 15.6 & 1.2  & 9.8  & 0.63 \\
		20 ppi thin  & 22.2	& 8.7& 1.4  &  3.9  & 0.44 \\
		20 ppi thick &26.9	& 10.2 & 0.4  &5.5 & 0.52 \\[0.1cm]
		60 ppi  & 16.5	& 5.3  & 0.42 & 1.5  & 0.28 \\[0.1cm]
		100 ppi & 11.9	& 3.2  & 0.25 & 0.49 & 0.15 \\
		\end{tabular}
		\caption{Fitted values for log-law parameters in dimensional and dimensionless form.  Listed values of $u_\tau/\kappa$, $y_d$, and $k_0$ are averages of the three estimates shown in figure~\ref{fig:log-law}, which were obtained for three different outer limits in the fitting procedure.  The displacement height is assumed to be zero for the smooth wall flow.}
		\label{tab:log-law}
	\end{center}
\end{table}

\subsection{Non-monotonic behaviour with pore size}\label{sec:nonmonotonic}
In many ways, the aforementioned transition from thick substrate behaviour, where turbulence penetration into the porous medium is limited and $y_d$ increases with permeability, to thin substrate behaviour, where turbulence penetrates the entire porous medium and $y_d \approx h$, is analogous to the $\lambda$-dependent transition from dense to sparse canopy behaviour proposed in the vegetated flow literature \citep{belcher2003adjustment,luhar2008interaction,nepf2012flow}.

As noted in the introduction, for vegetated flows the distance to which the flow penetrates into the canopy is dependent on the drag length-scale $(C_Da)^{-1}$, where $C_D$ is a representative drag coefficient and $a$ is the frontal area per unit volume.  As a result, the ratio of shear penetration to canopy height, $h$, is given by the dimensionless parameter $C_D a h = C_D \lambda$, where $\lambda$ is the solidity as before.  For dense canopies with $C_D \lambda \ge O(1)$, the shear layer does not penetrate the entire canopy and so an inflection point is expected in the mean profile. This gives rise to large-scale structures resembling Kelvin-Helmholtz vortices.  However, this instability mechanism is expected to weaken in sparse canopies.  For $C_D \lambda < O(0.1)$, turbulence penetrates the entire canopy and there is no inflection point in the mean profile \citep{nepf2012flow}.

The non-monotonic behaviour in mean velocity and turbulence intensity observed for the 10 ppi foam in the present experiments could be attributed to a similar weakening of the shear layer instability as the turbulence penetrates the entire porous medium, i.e. as $y_d \approx h$.  Consistent with this hypothesis, the reduced magnitude of the outer peak in streamwise intensity for the 10 ppi foam relative to the 20 ppi foam (figure~\ref{fig:ppi-u}b) suggests that the large-scale structures resembling Kelvin-Helmholtz vortices are weaker over the 10 ppi foam.  Since these structures contribute substantially to vertical momentum transfer, a reduction in their strength also translates into a smaller mean velocity deficit (figure~\ref{fig:ppi-u}a).

Note that there is evidence of non-monotonic behaviour as a function of solidity $\lambda$ in the rough-wall literature as well.  Based on a compilation of experimental data, \citet{jimenez2004turbulent} showed that the normalized roughness height (i.e. ratio of $k_0$ to roughness dimension) depends on the solidity $\lambda$, and that there are two regimes of behaviour.  For sparse roughness with solidity less than $\lambda \approx 0.15$, the normalized roughness increases with increasing $\lambda$.  In other words, an increase in frontal area leads to an increase in roughness drag.  However, for densely packed roughness with $\lambda \gtrsim 0.15$, the normalized roughness decreases with increasing $\lambda$ because the roughness elements shelter each other. Assuming that the relevant vertical dimension for turbulent flows over porous media is the displacement height $y_d$, we may expect similar non-monotonic behaviour for the normalized roughness $k_0/y_d = f(\lambda)$. The values for $k_0/y_d$ listed in table~\ref{tab:log-law} provide some support for this hypothesis: the normalized roughness height increases from $k_0/y_d = 0.15$ for the 100 ppi foam to $k_0/y_d = 0.63$ for the baseline 20 ppi foam, before decreasing to $k_0/y_d = 0.55$ for the 10 ppi foam. Bear in mind that, for identical $h$, the solidity is expected to increase with decreasing pore size from the `sparsely packed' 10 ppi foam to the `densely packed' 100 ppi foam.

\red{Although the solidity is a difficult parameter to measure for porous media, it may be estimated for the foams employed here based on simple geometric assumptions. Consider, for instance, a cubic lattice comprising thin rectangular ligaments of cross-section $d \times d$ and length $s$ (i.e., the pore size). Each unit cell of volume $s^3$ in this lattice comprises three orthogonal filaments aligned in the $x,y,$ and $z$ directions intersecting in a three-dimensional cross. Neglecting the overlapping volume at the center of the cross, the porosity is approximately $\epsilon \approx 1 - 3d^2s/s^3 = 1-3(d/s)^2$ for this geometry. For the foams tested here, the porosity is constant, $\epsilon \approx 0.97$. So, the above equation implies that the lattice must be geometrically similar with $d/s \approx \sqrt{(1-\epsilon)/3} = 0.1$. In other words, the ligament width $d$ increases linearly with pore size $s$ to maintain constant $\epsilon$. For this assumed geometry, the frontal area per unit volume for flow in either the $x$, $y$, or $z$ directions is $a \approx 2ds/s^3 = 2d/s^2$, since there are always 2 ligaments with area $ds$ normal to the flow. This results in solidity $\lambda = ah \approx 2(d/s)(h/s) = 0.2(h/s)$. This estimate suggests that the pore size threshold above which non-monotonic behavior is observed in the log-law constants ($s/h \ge 0.12$ in figure~\ref{fig:log-law}) corresponds to solidity $\lambda \sim O(1)$. Of course, since the foam pore structures do not resemble a cubic lattice (figure~\ref{fig:foam}), the numerical factors appearing in the equations above are unlikely to be accurate. However, the linear relationship between $\lambda$ and $h/s$ is expected to hold.

Finally, keep in mind that the non-monotonic behavior with solidity and pore size described above does not preclude the possibility of monotonic behavior with permeability Reynolds number $Re_k=u_\tau\sqrt{k}/\nu$. Although $y_d$ and $k_0$ are shown to decrease with increasing pore size, and hence permeability, from the 20 ppi foam to the 10 ppi foam, this decrease in the displacement and roughness heights is also accompanied by a decrease in $u_\tau/\kappa$  (Table~\ref{tab:log-law}). In other words, $Re_k$ may be lower for the 10 ppi foam compared to 20 ppi foam, even if $\sqrt{k}$ is higher.  Unfortunately, this hypothesis cannot be tested further without independent estimates of $u_\tau$.}

\subsection{A note on scaling}\label{sec:scaling}
Previous studies on turbulent flows over porous media indicate that shear penetration into the porous medium depends on the permeability length scale $\sqrt{k}$, which determines the effective flow resistance within the porous medium per the well-known Darcy-Forchheimer equation \citep{breugem2006influence,suga2010effects}.  This is also evident in the empirical relationship shown in equation (\ref{eq:ys}), which indicates that $y_d \approx 15.1\sqrt{k}$ for sufficiently high permeability Reynolds number $Re_k \gg 1$.  Although the permeability is related to geometric parameters such as pore size \citep[e.g. $\sqrt{k}/s \approx 0.08$ for the foams tested by][]{suga2010effects} and frontal area per unit volume, it is essentially a dynamic parameter that is typically estimated from fitting the Darcy-Forchheimer law:
\begin{equation}
-\frac{1}{\rho}\frac{\Delta P}{\Delta x} = \frac{\nu U_v}{k} + \frac{C_f}{\sqrt{k}}U_v^2,
\end{equation}
to experimental pressure drop measurements ($\Delta P/ \Delta x$) across porous media.  In the equation above, $U_v$ is the volume-averaged velocity and $C_f$ is the Forchheimer coefficient.  Since such pressure drop measurements are usually carried out at low Reynolds number in steady pipe or channel flow with uniformly distributed porous media (essentially a one-dimensional system), there are inherent risks in employing the resulting permeability values for unsteady, three-dimensional, spatially varying flows at higher Reynolds number. Further, the non-linear Forchheimer term that becomes increasingly important at higher speeds requires an additional coefficient $C_f$ that is not a universal constant.

To avoid these issues, the present study presents results primarily as a function of the normalized pore size, $s/h$. For example, figure~\ref{fig:log-law}b suggests that $y_d \propto s$ until it becomes comparable to foam thickness.  Another alternative would be to use the frontal area per unit volume, $a$, and solidity, $\lambda$, as the relevant scales. \red{The frontal area per unit volume is a difficult quantity to measure for complex porous media. However, the discussion presented in the previous section suggests that the solidity $\lambda$ increases linearly with $(h/s)$ for geometrically-similar porous media with constant porosity. As a result, the use of $s/h$ for scaling purposes also allows for greater reconciliation with the canopy flow and rough wall literature.}

\section{Conclusions}\label{sec:conclusions}
The experimental results presented in this paper show that turbulent boundary layers over high-porosity foams are modified substantially compared to canonical smooth wall flows. Development data in \S\ref{sec:development} suggest that the boundary layer adjusts relatively quickly to the presence of the porous substrate. Specifically, \red{for most of the foams tested,} the mean velocity profile adjusts to a new equilibrium over a streamwise distance $<10\delta$, which is similar to the adjustment length observed in previous literature for the transition from smooth to rough walls.\red{However, this rough-wall analogy does not hold for the thickest foam tested, which suggests that the foam thickness may also provide a bound on development length.} Fully-developed mean velocity profiles presented in \S\ref{sec:poresize} show the presence of substantial slip velocity ($>0.3U_e$) that is relatively insensitive to pore size \red{for foams of constant thickness. Profiles in \S\ref{sec:thickness} also show a near constant slip velocity over substrates with constant pore size and varying thickness.} These observations remain to be explained fully.

Profiles of streamwise intensity show the emergence of an outer peak at $y/\delta \approx 0.1$ over the porous substrates, which is associated with large-scale structures of length $2\delta - 4\delta$.  Such structures have also been observed in previous simulations and experiments, and are thought to arise from a Kelvin-Helmholtz instability associated with an inflection point in the mean profile. Although the magnitude of the outer peak in streamwise intensity generally increases with pore size, there is some evidence of weakening for the foam with the largest pore size. The log-law fits presented in \S\ref{sec:log-shift} provide further insight into this non-monotonic behaviour.  Specifically, the displacement height increases with \red{normalized pore size, $s/h$,} until it becomes comparable to the foam thickness.  Further increases in pore size beyond this point do not lead to an increase in $y_d$.  In other words, there is a transition from thick substrate behaviour, in which the thickness of the porous medium interacting with the flow is determined by pore size ($y_d \propto s$), to thin substrate behaviour, in which the flow penetrates the entire porous medium ($y_d \approx h$). The weakening in the outer layer structures may be attributed to this transition from thick to thin substrate behaviour.  Drawing an analogy to sparse canopy behaviour for vegetated flows, at the thin substrate limit, the mean velocity profile becomes fuller with increasing pore size and ultimately loses the inflection point. This results in a weakening of the Kelvin-Helmholtz instability. \red{For canopy flows, the transition from dense-canopy behavior to sparse-canopy behavior occurs as the solidity parameter becomes small, $\lambda \ll 1$. Simple geometric arguments show that $\lambda \propto h/s$ for the foams tested here, and that the transition from thick- to thin-substrate behavior occurs around $\lambda \sim O(1)$.}

Interestingly, the skewness of the near-wall velocity measurements increases substantially over the porous substrates relative to smooth wall values.  Further, this increase in skewness is correlated with an increase in the magnitude of the outer peak in streamwise intensity.  Given the link between skewness and the amplitude modulation phenomenon, these observations suggest that the large-scale structures that are energetic over porous media may have a modulating influence on the interfacial turbulence.  This is analogous to the interaction between VLSMs and near-wall turbulence in smooth wall flows at high Reynolds number.  Unfortunately, the near-wall velocity measurements collected as part of this study were not time resolved, and so did not allow for a quantitative evaluation of this effect. However, given the substantial similarities between turbulent flows over porous media and vegetation or urban canopies, further studies into such scale interactions could lead to the development of promising wall models for a variety of flows.

\section*{Acknowledgments}
This work was supported by the Air Force Office of Scientific Research under AFOSR grant No. FA9550-17-1-0142 (Program Manager: Dr. Douglas Smith).

\bibliographystyle{jfm}
\bibliography{Efstathiou2017}

\begin{thebibliography}{53}
\expandafter\ifx\csname natexlab\endcsname\relax\def\natexlab#1{#1}\fi
\def\au#1{#1} \def\ed#1{#1} \def\yr#1{#1}\def\at#1{#1}\def\jt#1{\textit{#1}}
  \def\bt#1{#1}\def\bvol#1{\textbf{#1}} \def\vol#1{#1} \def\pg#1{#1}
  \def\publ#1{#1}\def\arxiv#1{#1}\def\org#1{#1}\def\st#1{\textit{#1}}

\bibitem[Antonia \& Luxton(1971)]{antonia1971response}
{\sc \au{Antonia, RA} \& \au{Luxton, RE}} \yr{1971}  \at{The response of a
  turbulent boundary layer to a step change in surface roughness part 1. smooth
  to rough}.  \jt{Journal of Fluid Mechanics}  \bvol{48}~(4),  \pg{721--761}.

\bibitem[Battiato(2012)]{battiato2012self}
{\sc \au{Battiato, Ilenia}} \yr{2012}  \at{Self-similarity in coupled
  {B}rinkman/{N}avier--stokes flows}.  \jt{Journal of Fluid Mechanics}
  \bvol{699},  \pg{94--114}.

\bibitem[Belcher {\em et~al.\/}(2003)Belcher, Jerram \&
  Hunt]{belcher2003adjustment}
{\sc \au{Belcher, SE}, \au{Jerram, N} \& \au{Hunt, JCR}} \yr{2003}
  \at{Adjustment of a turbulent boundary layer to a canopy of roughness
  elements}.  \jt{Journal of Fluid Mechanics}  \bvol{488},  \pg{369--398}.

\bibitem[Breugem {\em et~al.\/}(2006)Breugem, Boersma \&
  Uittenbogaard]{breugem2006influence}
{\sc \au{Breugem, WP}, \au{Boersma, BJ} \& \au{Uittenbogaard, RE}} \yr{2006}
  \at{The influence of wall permeability on turbulent channel flow}.
  \jt{Journal of Fluid Mechanics}  \bvol{562},  \pg{35--72}.

\bibitem[Chandesris {\em et~al.\/}(2013)Chandesris, d'Hueppe, Mathieu, Jamet \&
  Goyeau]{chandesris2013direct}
{\sc \au{Chandesris, Marion}, \au{d'Hueppe, A}, \au{Mathieu, Benoit},
  \au{Jamet, Didier} \& \au{Goyeau, Benoit}} \yr{2013}  \at{Direct numerical
  simulation of turbulent heat transfer in a fluid-porous domain}.  \jt{Physics
  of Fluids}  \bvol{25}~(12),  \pg{125110}.

\bibitem[De~Graaff \& Eaton(2000)]{de2000reynolds}
{\sc \au{De~Graaff, David~B} \& \au{Eaton, John~K}} \yr{2000}
  \at{Reynolds-number scaling of the flat-plate turbulent boundary layer}.
  \jt{Journal of Fluid Mechanics}  \bvol{422},  \pg{319--346}.

\bibitem[DeGraaff \& Eaton(2001)]{degraaff2001high}
{\sc \au{DeGraaff, DB} \& \au{Eaton, JK}} \yr{2001}  \at{A high-resolution
  laser doppler anemometer: design, qualification, and uncertainty}.
  \jt{Experiments in fluids}  \bvol{30}~(5),  \pg{522--530}.

\bibitem[Detert {\em et~al.\/}(2010)Detert, Nikora \&
  Jirka]{detert2010synoptic}
{\sc \au{Detert, M}, \au{Nikora, V} \& \au{Jirka, GH}} \yr{2010}  \at{Synoptic
  velocity and pressure fields at the water--sediment interface of streambeds}.
   \jt{Journal of Fluid Mechanics}  \bvol{660},  \pg{55--86}.

\bibitem[Durst {\em et~al.\/}(1976)Durst, Melling \&
  Whitelaw]{durst1976principles}
{\sc \au{Durst, Franz}, \au{Melling, Adrian} \& \au{Whitelaw, James~H}}
  \yr{1976}  \at{Principles and practice of laser-doppler anemometry}.
  \jt{NASA STI/Recon Technical Report A}  \bvol{76}.

\bibitem[Duvvuri \& McKeon(2015)]{duvvuri2015triadic}
{\sc \au{Duvvuri, Subrahmanyam} \& \au{McKeon, Beverley~J}} \yr{2015}
  \at{Triadic scale interactions in a turbulent boundary layer}.  \jt{Journal
  of Fluid Mechanics}  \bvol{767},  \pg{R4}.

\bibitem[Favier {\em et~al.\/}(2009)Favier, Dauptain, Basso \&
  Bottaro]{favier2009passive}
{\sc \au{Favier, Julien}, \au{Dauptain, Antoine}, \au{Basso, Davide} \&
  \au{Bottaro, Alessandro}} \yr{2009}  \at{Passive separation control using a
  self-adaptive hairy coating}.  \jt{Journal of Fluid Mechanics}  \bvol{627},
  \pg{451--483}.

\bibitem[Finnigan(2000)]{finnigan2000turbulence}
{\sc \au{Finnigan, John}} \yr{2000}  \at{Turbulence in plant canopies}.
  \jt{Annual Review of Fluid Mechanics}  \bvol{32}~(1),  \pg{519--571}.

\bibitem[Ghisalberti(2009)]{ghisalberti2009obstructed}
{\sc \au{Ghisalberti, Marco}} \yr{2009}  \at{Obstructed shear flows:
  similarities across systems and scales}.  \jt{Journal of Fluid Mechanics}
  \bvol{641},  \pg{51--61}.

\bibitem[Herrin \& Dutton(1993)]{herrin1993investigation}
{\sc \au{Herrin, JL} \& \au{Dutton, JC}} \yr{1993}  \at{An investigation of
  {LDV} velocity bias correction techniques for high-speed separated flows}.
  \jt{Experiments in fluids}  \bvol{15}~(4),  \pg{354--363}.

\bibitem[Ho \& Huerre(1984)]{ho1984perturbed}
{\sc \au{Ho, Chih-Ming} \& \au{Huerre, Patrick}} \yr{1984}  \at{Perturbed free
  shear layers}.  \jt{Annual Review of Fluid Mechanics}  \bvol{16}~(1),
  \pg{365--422}.

\bibitem[Hutchins \& Marusic(2007)]{hutchins2007evidence}
{\sc \au{Hutchins, N} \& \au{Marusic, Ivan}} \yr{2007}  \at{Evidence of very
  long meandering features in the logarithmic region of turbulent boundary
  layers}.  \jt{Journal of Fluid Mechanics}  \bvol{579},  \pg{1--28}.

\bibitem[Itoh {\em et~al.\/}(2006)Itoh, Tamano, Iguchi, Yokota, Akino, Hino \&
  Kubo]{itoh2006turbulent}
{\sc \au{Itoh, Motoyuki}, \au{Tamano, Shinji}, \au{Iguchi, Ryo}, \au{Yokota,
  Kazuhiko}, \au{Akino, Norio}, \au{Hino, Ryutaro} \& \au{Kubo, Shinji}}
  \yr{2006}  \at{Turbulent drag reduction by the seal fur surface}.
  \jt{Physics of Fluids}  \bvol{18}~(6),  \pg{065102}.

\bibitem[Jackson(1981)]{jackson1981displacement}
{\sc \au{Jackson, PS}} \yr{1981}  \at{On the displacement height in the
  logarithmic velocity profile}.  \jt{Journal of Fluid Mechanics}  \bvol{111},
  \pg{15--25}.

\bibitem[Jacobi \& McKeon(2011)]{jacobi2011new}
{\sc \au{Jacobi, I} \& \au{McKeon, BJ}} \yr{2011}  \at{New perspectives on the
  impulsive roughness-perturbation of a turbulent boundary layer}.  \jt{Journal
  of Fluid Mechanics}  \bvol{677},  \pg{179--203}.

\bibitem[Jaworski \& Peake(2013)]{jaworski2013aerodynamic}
{\sc \au{Jaworski, Justin~W} \& \au{Peake, N}} \yr{2013}  \at{Aerodynamic noise
  from a poroelastic edge with implications for the silent flight of owls}.
  \jt{Journal of Fluid Mechanics}  \bvol{723},  \pg{456--479}.

\bibitem[Jimenez(2004)]{jimenez2004turbulent}
{\sc \au{Jimenez, Javier}} \yr{2004}  \at{Turbulent flows over rough walls}.
  \jt{Annual Review of Fluid Mechanics}  \bvol{36},  \pg{173--196}.

\bibitem[Jimenez {\em et~al.\/}(2001)Jimenez, Uhlmann, Pinelli \&
  Kawahara]{jimenez2001turbulent}
{\sc \au{Jimenez, Javier}, \au{Uhlmann, Markus}, \au{Pinelli, Alfredo} \&
  \au{Kawahara, Genta}} \yr{2001}  \at{Turbulent shear flow over active and
  passive porous surfaces}.  \jt{Journal of Fluid Mechanics}  \bvol{442},
  \pg{89--117}.

\bibitem[Jovic \& Driver(1994)]{jovic1994backward}
{\sc \au{Jovic, Srba} \& \au{Driver, David~M}} \yr{1994}  \at{Backward-facing
  step measurements at low reynolds number, $re_h$= 5000}.  \jt{NASA STI/Recon
  Technical Report N}  \bvol{94}.

\bibitem[Kim {\em et~al.\/}(2016)Kim, Blois, Best \&
  Christensen]{kim2016experimental}
{\sc \au{Kim, Taehoon}, \au{Blois, Gianluca}, \au{Best, Jim} \&
  \au{Christensen, Kenneth~T}} \yr{2016}  \at{Experimental study of a
  coarse-gravel river bed: Elucidating the nearwall and pore-space turbulent
  flow physics}.  \bt{In {\em River Flow 2016\/}},  \pg{pp. 950--955}.
  \publ{CRC Press}.

\bibitem[Kong \& Schetz(1982)]{kong1982turbulent}
{\sc \au{Kong, F} \& \au{Schetz, J}} \yr{1982} Turbulent boundary layer over
  porous surfaces with different surface geometries.  \bt{In {\em 20th
  Aerospace Sciences Meeting\/}},  \pg{p.~30}.

\bibitem[Kuwata \& Suga(2016)]{kuwata2016lattice}
{\sc \au{Kuwata, Y.} \& \au{Suga, K.}} \yr{2016}  \at{Lattice-{B}oltzmann
  direct numerical simulation of interface turbulence over porous and rough
  walls}.  \jt{International Journal of Heat and Fluid Flow}  \bvol{61}~(Part
  A),  \pg{145 -- 157}, tSFP9 Special Issue.

\bibitem[Kuwata \& Suga(2017)]{kuwata2017direct}
{\sc \au{Kuwata, Y} \& \au{Suga, K}} \yr{2017}  \at{Direct numerical simulation
  of turbulence over anisotropic porous media}.  \jt{Journal of Fluid
  Mechanics}  \bvol{831},  \pg{41--71}.

\bibitem[Le {\em et~al.\/}(1997)Le, Moin \& Kim]{le1997direct}
{\sc \au{Le, Hung}, \au{Moin, Parviz} \& \au{Kim, John}} \yr{1997}  \at{Direct
  numerical simulation of turbulent flow over a backward-facing step}.
  \jt{Journal of fluid mechanics}  \bvol{330},  \pg{349--374}.

\bibitem[Luhar {\em et~al.\/}(2008)Luhar, Rominger \&
  Nepf]{luhar2008interaction}
{\sc \au{Luhar, Mitul}, \au{Rominger, Jeffrey} \& \au{Nepf, Heidi}} \yr{2008}
  \at{Interaction between flow, transport and vegetation spatial structure}.
  \jt{Environmental Fluid Mechanics}  \bvol{8}~(5),  \pg{423--439}.

\bibitem[Mahjoob \& Vafai(2008)]{mahjoob2008synthesis}
{\sc \au{Mahjoob, Shadi} \& \au{Vafai, Kambiz}} \yr{2008}  \at{A synthesis of
  fluid and thermal transport models for metal foam heat exchangers}.
  \jt{International Journal of Heat and Mass Transfer}  \bvol{51}~(15),
  \pg{3701--3711}.

\bibitem[Manes {\em et~al.\/}(2011)Manes, Poggi \& Ridolfi]{manes2011turbulent}
{\sc \au{Manes, Costantino}, \au{Poggi, Davide} \& \au{Ridolfi, Luca}}
  \yr{2011}  \at{Turbulent boundary layers over permeable walls: scaling and
  near-wall structure}.  \jt{Journal of Fluid Mechanics}  \bvol{687},
  \pg{141--170}.

\bibitem[Marusic {\em et~al.\/}(2010)Marusic, Mathis \&
  Hutchins]{marusic2010predictive}
{\sc \au{Marusic, I}, \au{Mathis, R} \& \au{Hutchins, N}} \yr{2010}
  \at{Predictive model for wall-bounded turbulent flow}.  \jt{Science}
  \bvol{329}~(5988),  \pg{193--196}.

\bibitem[Marusic {\em et~al.\/}(2013)Marusic, Monty, Hultmark \&
  Smits]{marusic2013logarithmic}
{\sc \au{Marusic, Ivan}, \au{Monty, Jason~P}, \au{Hultmark, Marcus} \&
  \au{Smits, Alexander~J}} \yr{2013}  \at{On the logarithmic region in wall
  turbulence}.  \jt{Journal of Fluid Mechanics}  \bvol{716},  \pg{R3}.

\bibitem[Mathis {\em et~al.\/}(2009)Mathis, Hutchins \&
  Marusic]{mathis2009large}
{\sc \au{Mathis, Romain}, \au{Hutchins, Nicholas} \& \au{Marusic, Ivan}}
  \yr{2009}  \at{Large-scale amplitude modulation of the small-scale structures
  in turbulent boundary layers}.  \jt{Journal of Fluid Mechanics}  \bvol{628},
  \pg{311--337}.

\bibitem[Mathis {\em et~al.\/}(2013)Mathis, Marusic, Chernyshenko \&
  Hutchins]{mathis2013estimating}
{\sc \au{Mathis, Romain}, \au{Marusic, Ivan}, \au{Chernyshenko, Sergei~I} \&
  \au{Hutchins, Nicholas}} \yr{2013}  \at{Estimating wall-shear-stress
  fluctuations given an outer region input}.  \jt{Journal of Fluid Mechanics}
  \bvol{715},  \pg{163}.

\bibitem[Mathis {\em et~al.\/}(2011)Mathis, Marusic, Hutchins \&
  Sreenivasan]{mathis2011relationship}
{\sc \au{Mathis, Romain}, \au{Marusic, Ivan}, \au{Hutchins, Nicholas} \&
  \au{Sreenivasan, KR}} \yr{2011}  \at{The relationship between the velocity
  skewness and the amplitude modulation of the small scale by the large scale
  in turbulent boundary layers}.  \jt{Physics of Fluids}  \bvol{23}~(12),
  \pg{121702}.

\bibitem[McLaughlin \& Tiederman(1973)]{mclaughlin1973biasing}
{\sc \au{McLaughlin, DK} \& \au{Tiederman, WG}} \yr{1973}  \at{Biasing
  correction for individual realization of laser anemometer measurements in
  turbulent flows}.  \jt{The Physics of Fluids}  \bvol{16}~(12),
  \pg{2082--2088}.

\bibitem[Motlagh \& Taghizadeh(2016)]{motlagh2016pod}
{\sc \au{Motlagh, Saber~Yekani} \& \au{Taghizadeh, Salar}} \yr{2016}  \at{{POD}
  analysis of low {R}eynolds turbulent porous channel flow}.  \jt{International
  Journal of Heat and Fluid Flow}  \bvol{61}~(Part B),  \pg{665 -- 676}.

\bibitem[Nepf(2012)]{nepf2012flow}
{\sc \au{Nepf, Heidi~M}} \yr{2012}  \at{Flow and transport in regions with
  aquatic vegetation}.  \jt{Annual Review of Fluid Mechanics}  \bvol{44},
  \pg{123--142}.

\bibitem[Parikh(2011)]{parikh2011passive}
{\sc \au{Parikh, Pradip~G}} \yr{2011} Passive removal of suction air for
  laminar flow control, and associated systems and methods. US Patent
  7,866,609.

\bibitem[Patel(1965)]{PatelPreston1965}
{\sc \au{Patel, V.~C.}} \yr{1965}  \at{Calibration of the preston tube and
  limitations on its use in pressure gradients}.  \jt{Journal of Fluid
  Mechanics}  \bvol{23}~(1),  \pg{185--208}.

\bibitem[Pathikonda \& Christensen(2017)]{pathikonda2017inner}
{\sc \au{Pathikonda, Gokul} \& \au{Christensen, Kenneth~T}} \yr{2017}
  \at{Inner--outer interactions in a turbulent boundary layer overlying complex
  roughness}.  \jt{Physical Review Fluids}  \bvol{2}~(4),  \pg{044603}.

\bibitem[Poggi {\em et~al.\/}(2004)Poggi, Porporato, Ridolfi, Albertson \&
  Katul]{poggi2004effect}
{\sc \au{Poggi, Davide}, \au{Porporato, Amilcare}, \au{Ridolfi, Luca},
  \au{Albertson, JD} \& \au{Katul, GG}} \yr{2004}  \at{The effect of vegetation
  density on canopy sub-layer turbulence}.  \jt{Boundary-Layer Meteorology}
  \bvol{111}~(3),  \pg{565--587}.

\bibitem[Robinson(1991)]{robinson1991coherent}
{\sc \au{Robinson, Stephen~K}} \yr{1991}  \at{Coherent motions in the turbulent
  boundary layer}.  \jt{Annual Review of Fluid Mechanics}  \bvol{23}~(1),
  \pg{601--639}.

\bibitem[Rosti {\em et~al.\/}(2015)Rosti, Cortelezzi \&
  Quadrio]{rosti2015direct}
{\sc \au{Rosti, Marco~E}, \au{Cortelezzi, Luca} \& \au{Quadrio, Maurizio}}
  \yr{2015}  \at{Direct numerical simulation of turbulent channel flow over
  porous walls}.  \jt{Journal of Fluid Mechanics}  \bvol{784},  \pg{396--442}.

\bibitem[Ruff \& Gelhar(1972)]{ruff1972turbulent}
{\sc \au{Ruff, JF} \& \au{Gelhar, LW}} \yr{1972}  \at{Turbulent shear flow in
  porous boundary}.  \jt{J. Engrg. Mech}  \bvol{504}~(98),  \pg{975}.

\bibitem[Schlatter \& {\"O}rl{\"u}(2010)]{schlatter2010quantifying}
{\sc \au{Schlatter, Philipp} \& \au{{\"O}rl{\"u}, Ramis}} \yr{2010}
  \at{Quantifying the interaction between large and small scales in
  wall-bounded turbulent flows: A note of caution}.  \jt{Physics of fluids}
  \bvol{22}~(5),  \pg{051704}.

\bibitem[Schoppa \& Hussain(2002)]{schoppa2002coherent}
{\sc \au{Schoppa, W} \& \au{Hussain, Fazle}} \yr{2002}  \at{Coherent structure
  generation in near-wall turbulence}.  \jt{Journal of fluid Mechanics}
  \bvol{453},  \pg{57--108}.

\bibitem[Schultz \& Flack(2007)]{schultz2007rough}
{\sc \au{Schultz, MP} \& \au{Flack, KA}} \yr{2007}  \at{The rough-wall
  turbulent boundary layer from the hydraulically smooth to the fully rough
  regime}.  \jt{Journal of Fluid Mechanics}  \bvol{580},  \pg{381--405}.

\bibitem[Smits {\em et~al.\/}(2011)Smits, McKeon \& Marusic]{smits2011high}
{\sc \au{Smits, Alexander~J}, \au{McKeon, Beverley~J} \& \au{Marusic, Ivan}}
  \yr{2011}  \at{High--{R}eynolds number wall turbulence}.  \jt{Annual Review
  of Fluid Mechanics}  \bvol{43},  \pg{353--375}.

\bibitem[Suga {\em et~al.\/}(2010)Suga, Matsumura, Ashitaka, Tominaga \&
  Kaneda]{suga2010effects}
{\sc \au{Suga, K}, \au{Matsumura, Y}, \au{Ashitaka, Y}, \au{Tominaga, S} \&
  \au{Kaneda, M}} \yr{2010}  \at{Effects of wall permeability on turbulence}.
  \jt{International Journal of Heat and Fluid Flow}  \bvol{31}~(6),
  \pg{974--984}.

\bibitem[White \& Nepf(2007)]{white2007shear}
{\sc \au{White, Brian~L} \& \au{Nepf, Heidi~M}} \yr{2007}  \at{Shear
  instability and coherent structures in shallow flow adjacent to a porous
  layer}.  \jt{Journal of Fluid Mechanics}  \bvol{593},  \pg{1--32}.

\bibitem[Zagni \& Smith(1976)]{zagni1976channel}
{\sc \au{Zagni, Anthony~FE} \& \au{Smith, Kenneth~VH}} \yr{1976}  \at{Channel
  flow over permeable beds of graded spheres}.  \jt{Journal of the Hydraulics
  Division}  \bvol{102}~(2),  \pg{207--222}.

\end{thebibliography}

\end{document}